\documentclass[manuscript]{aastex}

\usepackage{longtable}
\usepackage{pdflscape}
\shorttitle {Statistical Classification of bulges}
\shortauthors{Tanuka Chattopadhyay \& Pradip Karmakar}

\begin{document}

\title{Cosmic history of integrated galactic stellar initial mass function : a simulation study }

\author{Tanuka Chattopadhyay,\altaffilmark{1}}
\affil{$^1$Department of Applied Mathematics, Calcutta
            University,
             92 A.P.C. Road, Calcutta 700009, India} \email{tanuka@iucaa.ernet.in}

\author{Tuli De,\altaffilmark{2}}
\affil{$^2$Department of Oncology, Novartis Health Care Pvt.
Ltd.,Hyderabad, India} \email{tuli.de@novartis.com }

\author{Bharat Warlu,\altaffilmark{3}}
\affil{$^3$ Zensar Technology, Pune } \email{bwarule@gmail.com\\
and }

\author{Asis Kumar Chattopadhyay\altaffilmark{4}}
\affil{$^4$Department of Statistics, Calcutta University, 35 B.C.
Road, Calcutta 700019, India} \email{akcstat@caluniv.ac.in  }

\begin{abstract}

\noindent  Theoretical and indirect observational evidences
suggest that stellar initial mass function (IMF) increases with
redshift. On the other hand star formation rates (SFR) may be as
high as 100 $M_{\odot}$ yr$^{-1}$ in star burst galaxies. These
may lead to formation of massive clusters hence massive stars to
make the integrated galactic stellar initial mass function (IGIMF)
top heavy (i.e. proportion of massive stars is higher than less
massive stars). We investigate the joint effect of evolving IMF
and several measures of SFR in dependence of galaxy wide IMF. The
resulting IGIMF have slopes $\alpha_{2,IGIMF}$ in the high mass
regime, which is highly dependent on the minimum mass of the
embedded cluster ($M_{ecl,min}$), star formation rates and mass
spectrum indices of embedded clusters (viz. $\beta$). It is found
that for z $\sim$ 0 - 2, $\alpha_{2,IGIMF}$ becomes steeper (i.e.
bottom heavy), for z $\sim$ 2 - 4, $\alpha_{2,IGIMF}$ becomes
flatter (i.e. top heavy ) and  from z $\sim$ 4 onwards
$\alpha_{2,IGIMF}$ becomes again steeper. The effects are faster
for higher values of $\beta$. $\alpha_{2,IGIMF}$ is flatter also
for higher values of $M_{ecl,min}$. All these effects might be
counted for the joint effect of increasing temperature of the
ambient medium as well as varying SFR with increasing redshift.

\end{abstract}

\section{Introduction}

\noindent The form of stellar initial mass function is of
considerable debate in the present era as it describes the nature
of stellar population, the ratio of high mass to low mass stars
and influences the dynamical evolution of star clusters as well as
star formation history of the whole galaxy. Usually it is derived
using observed luminosity function together with an assumed
mass-to-light ratio for the stars under consideration. Generally,
IMFs, as suggested by various authors, are either of Salpeter type
(Salpeter 1955) or consists of segmented power laws (Scalo 1986)
or of lognormal type (Chabrier 2003) with a turnover at some
characteristic mass $m_{c}$. The power law slope at high masses is
probably close to the Salpeter value (Salpeter 1955) with
$\frac{dN}{dm}\propto m^{-\alpha}$, $\alpha=2.35$ with an
uncertainty $\sim$ 0.3 (Chabrier 2003).  There are conflicting
 views or evidences of universality of IMF at present time e.g.
there are evidences of mass segregation to some extent in young
massive clusters (Zwart et al. 2010 and references therein) and
some variations in star burst galaxies (Gunawardhana et al. 2011).
On the contrary there are bottom heavy IMF in massive ellipticals
(Ferreras et al. 2013) and no direct evidence for rapid variation
of IMF within Milky Way disc (Kroupa 2001; Chabrier 2003). Again
there is possibility of variation of IMF with time (hence
redshift), metallicity and environment. In a work Larson (Larson
2003; Larson 2005) has argued that the characteristic mass is
primarily determined by Jeans mass which depends on the
temperature. Hence, with the increase of temperature (cosmic
microwave background temperature was higher at higher redshift),
one might expect that low mass star formation is disfavored
resulting in a top heavy IMF with a temperature scaling with
redshift as (1 + z). Therefore at sufficiently high redshift, mass
scales as $(1 + z)^{\frac{3}{2}}$, increasing the fraction of high
mass stars. Larson (2005) has suggested that at z = 5 the
characteristic mass may be higher than present day's value
by an order of magnitude.\\

\noindent Recently from various observations it is clear that
stars form in embedded clusters (Lada \& Lada 2003; Kroupa et al.
2005). These clusters also follow a mass function which is again a
power law, $\xi_{ecl} (M) \propto M_{ecl}^{-\beta}$. This is known
as embedded cluster mass function (hereafter ECMF). The maximum
mass of the ECMF, $M_{ecl,max}$ has been found to depend on the
star formation rate of the galaxy (Weidner \& Kroupa 2004; Bastian
2008) and is given as
\begin{equation}
\label{Meclmax,SFR}
 M_{ecl,max}=8.5\times10^{4}\times SFR^{0.75}
\end{equation}

\noindent  where SFR is in $M_{\odot} yr^{-1}$. Weidner et al.
(2013) have suggested a time dependent IMF for elliptical galaxies
to account for an excess of low mass stars in these galaxies. They
modeled the SFR as a function of time. The SFR reaches a maximum
initially and then asymptotically reduces to zero with time. {\bf
They have discussed a two stage star formation scenario in giant
elliptical galaxies and have given an alternative hypothesis over
time independent bottom heavy IMF in these galaxies. They have
proposed that initially there is a strong star burst stage with
top heavy IMF and it is followed by a prolonged stage with a
bottom heavy IMF. The latter result originates from many low mass
clouds (i.e. a high value of beta) formed as a result of
fragmentation of the gaseous component.} Similar trend of SFR has
also been found for various elliptical galaxies as a function of
redshift (Spaans \& Carollo 1997). Various theoretical models for
star formation are difficult to testify as current observational
result from cosmological studies do not measure IMF slopes and
SFRs for individual galaxy but study indirect evidence for whole
populations and average the
results over the galaxy luminosity function.\\

\noindent  Various authors have suggested a changing IMF with
redshift (van Dokkum 2008; Larson 1998, 2005). Some have dealt
with the resulting form of IGIMF using various empirical measures
of SFR, minimum mass of embedded clusters (Weidner et al. 2010,
2013). But no studies have not been made so far to investigate the
joint effect of the time varying IMF together with a time varying
SFR derived from an observed SFR function varying indirectly with
redshift on the resulting slopes of IGIMF. Hence our aim is to
study the cosmic history of galaxy wide IMF in this concern
for varying IMF as well as SFR. \\

\noindent Now for the SFR in equation (1) there are no
observational values available so far for SFR directly as a
function of redshift which helps to study the cosmic star
formation history. Smit et al. (2012) have computed the SFR
function $\phi(SFR)$ (in $Mpc^{-3} dex^{-1}$, which is a Schecter
function (Schecter 1976)), using a characteristic SFR parameter
denoted by $SFR^*$, whose values are given from published studies
for redshifts z $\sim$ 4 - 7. The other values for z $\sim$ 0 - 3
are taken from literature (Bothwell et al. 2011; Bell et al. 2007;
Sobral et al. 2012; Magnelli et al. 2011 and Reddy et al. 2008).
But $SFR^*$ characterizes a particular value of SFR varying with
redshift. This excludes for many others for z, a few with higher
SFR than $SFR^*$, many with a low SFR than $SFR^*$. To cover up
the above mentioned uncertainty we have computed also the quartile
values of SFR (viz. SFR1, SFR2, SFR3) from SFR function by
converting it to a probability density function. Thus through this
process galaxies with low, intermediate and high SFR are also
being involved and we have an overall view of the variation of
cosmic star formation history of galaxies. In this respect Smit et
al. (2012) results are adequate for this study. {\bf In this
regard it is to be mentioned that there are conflicting views of
star formation histories of high red shift galaxies ($z\sim 2-7$).
Reddy et al.(2012) have computed the SFR of high red shift
galaxies in the range $z \sim 2-7 $ from  spectral energy
distribution (SED) as well as from infra red and  ultra violet
imaging (IR+UV). It shows an exponential decrease with different
slopes  of SFR with time. Duncan et al (2014) have also computed
the star formation histories for high red shift galaxies using
data in the CANDELS GOODS South field and their SFR density
estimates are higher than previously observed in this regime. It
will be interesting to incorporate all these studies and compare
the resulting IGIMF in a future study.}
\\

\noindent In the present problem we have replaced the SFR in
equation (1) by several significant measures of SFR, e.g. SFR1,
SFR2, SFR3 and SFR$^*$ which correspond to  the first, second,
third quartiles and characteristic value of $log_{10}$SFR
distribution computed by Smit et al. (2012). The first three
measures have been computed from the normalized $log_{10}$SFR
distribution (Smit et al. 2012) as a function of z and the last
one is given in Smit et al. (2012) as a function of z. All the
four values satisfy tapered power law function (viz. subsection
2.1). As a result $M_{ecl.max}$ becomes indirectly a function of z
in all four cases. The significance for the use of the above three
quartiles and characteristic star formation rate  and their
derivation have been discussed
in detail in sub section 2.1. \\

\noindent In the work by Chattopadhyay et al. (2011), the authors
have considered the random fragmentation of young massive clusters
in our Galaxy as well as in external galaxies. There they found no
correlation between the maximum mass of a star to its embedded
cluster mass. Existence of a correlation between the above two
affects the star formation history of the parent cloud. Low mass
clouds do not have enough mass to form high mass stars (Bruzual\&
Charlot 2003; Larson 2006; Weidner et al. 2007; Weidner et al.
2010). Formation of massive stars is possible if they accrete
their masses from surrounding. On the other extent, in massive
clouds, once the high mass stars are formed, their ionizing
radiation removes remaining gas (Weidner et al. 2006). This stops
formation of low mass stars. This fact is reflected in some
observations (Weidner et al.
2010 and references therein).\\

\noindent In contrast, the recent observations by Maschberger\&
Clarke (2008) and Parker \& Goodwin (2007) included several
examples of low mass clusters containing high mass stars. Corbelli
et al. (2010) found that for YMCs of M33, such strict correlation
does not exist. Moreover unresolved binaries play an important
role. Elmegreen (2006) argued that clusters are built
stochastically: the large amount of molecular gas present in the
star formation region allows high mass stars to form even in a low
SFR region i.e, the entire range of masses (0.01$M_{\bigodot}$ to
150$M_{\bigodot}$) is possible even in a low SFR region. Andrews
et al. (2013) studied the dwarf star burst galaxy and found no
such correlation between maximum mass of star with cluster mass.
Furthermore previous observations included very small numbers of
YMCs  ($\sim 10^{5}-10^{6}M_{\bigodot}$) (viz. $\sim 10\%$ Weidner
et al.2013), for which any such correlation is difficult to
predict.\\

\noindent $Cervi\tilde{n}o$ et al. (2013) have argued that
simulated sampling is not in contradiction of a possible
$m_{max}-M_{ecl}$ correlation and it depends on the star formation
process and the assumed definition of stellar cluster. Hence
considering all aspects, we have not assumed any such correlation
but only the scenario that massive clouds have a general tendency
to form massive stars and have taken the minimum and maximum mass
of stars to be 0.1$M_{\bigodot}$ (Hass \& Ander 2010) and
150$M_{\bigodot}$ respectively. \\

\noindent In the present work, we have considered the resulting
integrated galactic stellar mass function (IGIMF) as a function of
redshift due to random fragmentation of embedded clusters of
various masses present in that parent galaxy. Section 2 describes
the model with model parameters. Section 3 describes the method.
Sections 4
and 5 give results and conclusion.\\

\section{The Model}

\noindent In the present model, the star formation scenario of a
galaxy has been considered. The component of a galaxy that forms
stars consists of molecular clouds and each cloud, under
gravitational instability undergo hierarchical fragmentation
(Hoyle 1953) giving rise to a number of fragments of various
masses. These fragments ultimately form stars and we get what we
call stellar initial mass function of these star clusters embedded
into the molecular clouds. For simplicity we have assumed that the
IMF, in each parent cloud has the same distributional form and
this is a segmented power law of the form

\begin{equation}
\noindent Let,  \  \xi_{IMF} (m) =\frac{dN}{dm}=   \left\{
      \begin{array}{lcl}
        Am^{-\alpha_{1,IMF}}; m_{min}<m\leq m_{c}\\
        Bm^{-\alpha_{2,IMF}}; m_{c}<m\leq m_{max}\\
              \end{array}
    \right.
\end{equation}

 \noindent where $m_{min}$ and $m_{max}$ are the minimum and maximum masses of
 the stars, $m_{c}$ is the characteristic mass at which the
 turnover occurs. The values of A and B are calculated as follows: \\
 \noindent Since $\xi _{IMF}$ is a probability density function,
 we have the normalization condition,
\begin{equation}
\int_{m_{min}}^{m_{max}} \xi_{IMF}(m) dm = 1
\end{equation}

\noindent Also the IMF is continuous at $m_c$. Hence the
continuity condition gives $Am_c^{-\alpha_{1,IMF}} = B
m_c^{-\alpha_{2,IMF}}$ i.e.

\begin{equation}
A = B m_c^{\alpha_{1,IMF} - \alpha_{2,IMF}}
\end{equation}

\noindent Then from equation(3)\\

$\int_{m_{min}}^{m_c} Am^{- \alpha_{1,IMF}} dm +
\int_{m_c}^{m_{max}} B m^{- \alpha_{2,IMF}} dm = 1$\\

\noindent Substituting the value of A from equation (4) we get, \\

$B = \frac{1}{ [ \frac{m_c^{\alpha_{1,IMF} - \alpha_{2,IMF}}}{(1 -
\alpha_{1,IMF})} ( m_c^{1-\alpha_{1,IMF}} -
m_{min}^{1-\alpha_{1,IMF}} ) +}$
\begin{equation}
{ \frac{1} {1-\alpha_{2,IMF}}( m_{max}^{1-\alpha_{2,IMF}} -
m_c^{1-\alpha_{1,IMF}} ) ] }
\end{equation}

\noindent Then using B from equation (5), A is found from equation
(4). The representative values of $\alpha_{1,IMF}$ and
$\alpha_{2,IMF}$ are chosen as 1.25 ( the maximum value is 1.25 in
low mass regime, Bastian et al. 2010) and 2.35
(Salpeter 1955) respectively.\\

 \noindent The values of $m_{min}$ and $m_{max}$ are chosen as
 0.1$M_{\bigodot}$ and 150$M_{\bigodot}$ respectively (Zinnecker\& York 2007).
The value of the characteristic mass at z = 0 is taken as $m_{c}$
$(z=0)$ = 0.3$M_{\bigodot}$ (Larson 2005). Since, we have assumed
a top heavy IMF with increasing redshift the characteristic mass
is given by

\begin{equation}
\label{mc z}
 m_{c}=D(1+z)^\frac{3}{2}
\end{equation}

\noindent where D is determined from the condition that at z = 0,
$m_{c}$ = 0.3$M_{\bigodot}$ (Larson2005).  The choice of the above
relation is not arbitrary but has a strong physical ground. The
influence of temperature on the Jeans mass (Jeans 1902) is a very
well known phenomenon. Larson (1998, 2005) has discussed that
characteristics turnover mass may be primarily determined by
thermal Jeans mass which is strongly influenced by temperature
($\sim T^{3/2}$) at fixed density. Hence it is expected that
environment, where heating occurs through far infrared radiation,
disfavors formation of low mass stars. Such extreme environments
really occur in the super clusters at the centre of Milky Way.
Some young super clusters at the centre of M82 really appear to
have a top heavy IMF  (e.g. Rieke et al.1993; McCrady et al. 2003)
along with those at the centre of our Galaxy (Stolte et al. 2005;
Maness et al. 2007). The mass functions in these super clusters
also have the additional effects of complex dynamical phenomena
which make them top heavy over time (McCrady et al. 2005; Kim et
al. 2006; Harayama et al. 2008). At the initial stage of star
formation in giant as well as in dwarf galaxies the star formation
occurs in 'burst' rather than through a continuous process
(Steidel et al. 1996; Blain et al. 1999; Lacey et al. 2008). This
means IMF becomes more and more top heavy at redshifts 1 - 3 and
beyond. Also at high redshift the metallicity was lower in star
forming clouds. Thus initially cooling process was not efficient
which may lead to an extremely top heavy IMF for the first
generation stars (Abel et al. 2002; Bromm et al. 2002). Hence IMF
may depend on redshift. Cosmic microwave background temperature
(CMB) plays a significant role for increasing the temperature of
the medium which scales as (1 + z). Beyond z $\sim$ 2, the CMB
temperature exceeds the temperature of the Galactic molecular
clouds (Evans et al. 2001; Tafalla et al. 2004). Hence it can be
speculated that the characterstic mass $ m_c \sim T^{3/2}$ at
fixed density, varies with redshift as $(1 + z )^{3/2}$, at high
redshift leading to a top heavy IMF (Larson 1998). The effect
becomes more pronounced when pressure is taken into account and
Larson (2005) has shown that at z = 5, $m_c$ becomes higher by an
order of magnitude than its present value. The direct evidence of
a top heavy IMF at high redshift is very rare though there are few
observations e.g. blue rest frame ultra violet colours of galaxies
at z $\sim$ 6 may imply a top heavy IMF (Stanway et al. 2005).
Tumlinson (2007) finds that the properties of carbon enhanced
metal poor stars in our Galaxy are best explained by relatively
large number of stars in the mass range 1
- 8 $M_{\odot}$ at high redshift. \\

\noindent The maximum mass of the embedded cluster $M_{ecl,max}$
has been assumed  to be a function of SFR and indirectly becomes a
function of redshift as discussed in Section 1 and equation (1).
The ECMF is assumed to be

\begin{equation}
\label{ECMF}
 \xi_{ecl}(M) = \frac{dN}{dM_{ecl}} = EM_{ecl}^{-\beta},
M_{ecl,min}\leq M_{ecl}\leq M_{ecl,max}
\end{equation}

\noindent where $M_{ecl,min}$ is the minimum mass of embedded
cluster. The value of index $\beta$ is around 2 (Zhang \& Fall
1999; de Grijs et al. 2003; McCraday \& Graham 2007). Some studies
also suggest flatter slopes like 1.8 (Dowell et al. 2008). The
mass spectrum of giant molecular clouds shows, $\beta\sim 1.7$
(Rosolowsky 2005). In the present work we have considered $\beta$
ranging from 2 to 2.6. The lower limit of embedded cluster is
considered as a parameter having values 500 and 1000
$M_{\bigodot}$ respectively. The value of the constant E in
equation (7) is determined assuming  galaxy mass of $5 \times
10^9$ $M_{\odot}$, $5 \times 10^{10}$ $M_{\odot}$ and $5 \times
10^{11}$ $M_{\odot}$ respectively as representative values of
dwarf, intermediate and giant galaxies, whose 30 percent mass has
been exhausted due to star formation (Lada et al. 1984;
Elmegreen \& Clemens 1985; Verschueren et al. 1982). \\

\noindent Then the integrated galactic initial mass function
(hereafter IGIMF) as a function of fragment mass m and redshift z,
is the collection of all IMFs of all the parent clusters (Kroupa
\& Weidner 2003; Weidner \& Kroupa 2005; Vanbeveren 1982 ) which
is,

\begin{equation}
\label{IGIMF}
 \xi_{IGIMF}(m,z)=\int_{M_{ecl,min}}^{M_{ecl,max}}
\xi_{IMF}(m) \ \xi_{ecl}(M) \,dM_{ecl}
\end{equation}
All the values of the parameters considered are listed in Table 1.

\subsection{Various measures of star formation rate}

\noindent To compute various measures of SFR as a function of
redshift (z) we start with the star formation rate function
$\phi$(SFR) {\bf (in $Mpc^{-3} dex ^{-1}$)} derived by Smit et
al.(2012) which is,

\begin{equation}
\phi(SFR) dSFR = \phi^{*}_{SFR} (\frac{SFR}{SFR^*})^{\alpha_{SFR}}
exp (- \frac{SFR}{SFR^*}) \frac{dSFR}{SFR^*}
\end{equation}

\noindent Where $\phi^*_{SFR} $, $\alpha_{SFR}$ and SFR$^*$ are
various Scechter parameters given as a function of z in Tables 2
and 3 of Smit et al. (2012).

\noindent Then $log_{10}\phi$(SFR) function is,

\noindent $log_{10}\phi(SFR) \ dlog_{10}SFR =[log
_{10}(\phi^{*}_{SFR}) +
{\alpha_{SFR}}  \ log _{10}(\frac{SFR}{SFR^*}) - $\\
\begin{equation}
 \frac{SFR}{SFR^*} log _{10} e  - log_{10}SFR^*] d log_{10} SFR
\end{equation}

\noindent At first step we convert $log _{10} \phi$(SFR) to a
density function at each z  dividing by ,\\

$ T = \int _0^{log _{10}SFR_{max}} log_{10} \phi(SFR) \  dlog(SFR)$,\\

\noindent where $log_{10}$SFR$_{max}$ is the maximum value of
$log_{10}$SFR at a particular z, taken from Fig.2 of Smit et al.
(2012) for z = 4, 5, 6, 7. For other values of z the values of
$log_{10}$SFR$_{max}$ were found by plotting the function. The
lower boundary is not strictly zero and it includes negative
values also (Fig. 2 of Smit et al. 2012) but the number of
observations for negative values of  $log _{10}$ SFR  decreases
and for z $\sim $ 6 , 7 (viz. Table 1 of Smit et al. 2012) it is
just one. So the negative part of $log _{10}$ SFR contribution is
insignificant compared to positive part on the basis of
observational range. Therefore we limited our study of $log _{10}$
SFR  from 0 to $log_{10}$SFR$_{max}$ due to lack of observational
points for the negative part and we have worked
with available observational range. \\

\noindent Then the c.d.f of $log_{10}$SFR distribution is given
by\\

\noindent $\frac{1}{T} \int _0^{log_{10}SFR} log_{10}\phi(SFR) \
dlog_{10}SFR $ = \\

$\frac{1}{T} \int_0^{log_{10}SFR} [log _{10}(\phi^{*}_{SFR}) +
{\alpha_{SFR}}  \ log _{10}(\frac{SFR}{SFR^*}) - $\\
\begin{equation}
 \frac{SFR}{SFR^*} log _{10} e  - log_{10}SFR^*] d log_{10} SFR
\end{equation}

\noindent In equation (11) when the LHS is 0.25 then the
corresponding value of $log_{10}$SFR is $log_{10}$SFR1 i.e. the
first quartile. This is a point such that at this point 75 \% of
the galaxies have $log_{10}$SFR $> log_{10}$SFR1 and remaining 25
\% of the galaxies have $log_{10}$SFR $\le log_{10}$SFR1 at a
particular z. Similarly we have values of $log_{10}$SFR1 for
different z for different values of $log_{10}\phi^*_{SFR}$,
$\alpha_{SFR}$ and $log_{10}$SFR$^*$ at different z given in
Tables 2 and 3 of Smit et al. (2012). We fit the values of
$log_{10}$SFR1 at different z by a tapered power law function of
the form {\bf $ log _{10} SFR1 \propto z^{- \gamma} [ 1 - e
^{(-z/\delta)^x}]$}, where $\gamma$, $\delta$, x are constants. We
repeat the above process for values of c.d.f as 0.5 and 0.75 and
we get $log_{10}$SFR2 and $log_{10}$SFR3 as function of z. The
fitted tapered power law functions against $log_{10}$SFR1,
$log_{10}$SFR2, $log_{10}$SFR3, $log_{10}$SFR$^*$ are shown in
Figs 1 - 4 along with their p-values.  We have not only fitted
tapered power law but also performed the goodness of fit test for
which the p-values are much higher (more than 0.25). Hence we can
accept the null hypothesis (the tapered power law is a suitable
curve). The significance of constructing these quartile points and
$log_{10}$ SFR$^*$ as function of z is that we will have a clear
view how SFR of most of the galaxies vary with redshift. Among all
SFR2 is the most representative measure because $log_{10}$SFR2 is
the median value of $log_{10}$SFR distribution.

\section{Method}

\noindent To generate a sample of embedded cluster masses from
power law with a given range of values it can be considered as
truncated Pareto distribution. For this we have used the standard
method of inverting the cumulative distribution function (cdf).
Let X be a random variable with probability density function (pdf)
f(x) and cdf F(x), where

\begin{equation}
\label{CDF}
 F(x) =\int_{-\infty}^{x} f(x)\,dx
\end{equation}

,\[\int_{-\infty}^{\infty} f(x)\,dx=1\].\\
\noindent We know that the cdf F(x) follows Uniform distribution
over the range (0,1). Hence a simulated value x of X can be
obtained by solving the equation F(x) = r, where r is a random
fraction. Thus one simulated value is given by x = $F^{-1}$(r).
Corresponding to n choices of r, we will have n values of x giving
a simulated sample of size n. Of course the above method is valid
when the inverse function of F exists, which is true in the
present case. To generate the value of X, it is necessary to know
the parameters of the Pareto distribution and those are the
constants in the Power laws already known from physical
considerations. In the present work, in equation (7) lower limit
of cluster  mass is taken as $M_{ecl,min}$ instead of $-\infty$,
with f(x) = $\xi_{ecl}(M)$
 for
sampling cluster masses.\\

\noindent The method of generating  samples of stellar masses from a
segmented power law (truncated Pareto distribution) is as follows:\\

\begin{equation}
\noindent Let,  \  \xi_{IMF} (m) =\frac{dN}{dm}=   \left\{
      \begin{array}{lcl}
        Am^{-\alpha_{1,IMF}}; m_{min}<m\leq m_{c}\\
        Bm^{-\alpha_{2,IMF}}; m_{c}<m\leq m_{max}\\
              \end{array}
    \right.
\end{equation}

\noindent where A and B are constants to be determined by
equations (4) and (5). Now, to generate samples from the above
power laws, we use a
conditional cdf defined as follows:\\

$F_{1}(m)=P(X<m|m_{min}<x<m_{c})=\frac{F(m)-F(m_{min})}{F(m_{c})-F(m_{min})}$
\begin{equation}
m_{min}<m<m_{c}.
\end{equation}
In the same way,\\
$F_{2}(m)=P(X<m|m_{c}<x<m_{max})=\frac{F(m)-F(m_{c})}{F(m_{max})-F(m_{c})}$
\begin{equation}
m_{c}<m<m_{max}
\end{equation}
\noindent We use the method of inversion to draw samples using
these two conditional cdfs. Firstly when $m_{min} < m < m_{c}$, we
draw a random sample say $u_{1}$ from a Uniform distribution,
i.e., U(0,1) and equate it to
\begin{equation}
F_{1}(m)=\frac{F(m)-F(m_{min})}{F(m_{c})-F(m_{min})}=u_{1}
\end{equation}
So that inverting it we get the expression for the sample m as
\begin{equation}
m=[u_{1}\times(m_{c}^{(1-\alpha_{1,IMF})}-m_{min}^{(1-\alpha_{1,IMF})})+m_{min}^{(1-\alpha_{1,IMF})}]^{\frac{1}{(1-\alpha_{1,IMF})}}
\end{equation}
Thus when $u_{1}=0, m=m_{min}$ and when $u_{1}=1, m=m_{c}$.\\
Similarly, when $m_{c} < m < m_{max}$, we draw a random sample say
$u_{2}$ from a Uniform distribution, i.e., U(0,1) and equate it to
\begin{equation}
F_{2}(m)=\frac{F(m)-F(m_{c})}{F(m_{max})-F(m_{c})}=u_{2}
\end{equation}
So that inverting it we get the expression for the sample m as
\begin{equation}
m=[u_{2}\times(m_{max}^{(1-\alpha_{2,IMF})}-m_{c}^{(1-\alpha_{2,IMF})})+m_{c}^{(1-\alpha_{2,IMF})}]^{\frac{1}{(1-\alpha_{2,IMF})}}
\end{equation}
Thus when $u_{2}=0, m=m_{c}$ and when $u_{2}=1, m=m_{max}$.\\

\noindent We simulate from $F_{1}(m)$ as long as the total mass of
the embedded cluster is equal to the mass in the low mass regime
(viz $m_{min}<m<m_{c}$) and then we simulate from $F_{2}(m)$ for
the high mass regime (viz. $m_{c}<m<m_{max}$). The mass fractions
for each embedded cluster in the low and high mass regimes are
computed at the beginning for different $m_c$s.

\noindent In the present work we have simulated random samples
from various segmented power law
distributions as follows.\\

\noindent (i) First we simulate a sample of embedded cluster
masses following the normalized power law given in equation (7),
where the maximum mass is computed at any particular z following
equation (1) for different SFR (viz. SFR1, SFR2, SFR3 and
SFR$^*$). The simulation is continued as long as the total mass of
the embedded cluster is less or equal to 30 percent of the total mass of the galaxy.\\

\noindent (ii) Secondly for each mass of a parent cluster we
simulate a sample of stellar masses following the segmented power
law (as discussed before) distributions, given in equation (2) at
a particular value of z, so that the value of the characteristic
mass, $m_{c}$ is prefixed at that value of z (refer to equation
(6)). Each time a stellar mass is simulated, the total mass of the
previous stellar masses is checked with the total mass of the
embedded cluster and as soon as it exceeds the
mass of the parent cluster, the simulation is stopped.\\

\noindent (iii) Finally the mass spectrum of all simulated stellar
masses from all the parent clusters of the galaxy is computed and
fitted by
segmented power laws, to give the resulting form of the IGIMF.\\

\noindent (iv) The above procedure is performed at various SFR
(viz. SFR1, SFR2, SFR3, SFR$^*$), redshifts z, $M_{ecl,min}$ and
$\beta$ respectively.

\section{Results and interpretations}

\noindent Tables 2 - 9 and Figs. 5 - 11 show the resulting IGIMF
slopes, for stars in a galaxy which consists of segmented power
laws with slopes, $\alpha_{1,IGIMF}$ in low mass regime and
$\alpha_{2,IGIMF}$ in high mass regime for various values of SFR,
$\beta$, $M_{ecl,min}$ and redshift z = 0.1 to z = 6.8
respectively.
The following observations are  envisaged.\\

\noindent (i) As z increases, $\alpha_{2,IGIMF}$ becomes
systematically rising up to $z \sim 2$ and then starts falling. It
is again rising from around $z\sim 4$ and runs down around $z\sim
6$. The effect is more pronounced for $\beta = $ 2 and 2.4. For
$\beta $ = 2.6 the rise and fall are comparatively small (viz.
Figs.5 - 7).  We have also tested for equality of means of
$\alpha_{2,IGIMF}$ values over $\beta$ and $M_{ecl,min}$ = 500
$M_{\odot}$ and 1000 $M_{\odot}$ respectively for SFR2 (e.g.) by
MANOVA test (Multivariate Anakysis for Variance) . The test has
been rejected in all cases (viz. p -values are 0.0278 and 0.0508
respectively which are very small). This might be explained as
follows. Though from z = 0 - 2 the SFR and hence $M_{ecl,max}$ are
increasing (viz. equation 1), due to low temperature of the
ambient medium Jeans mass does not favor formation of massive
stars. That is why $\alpha_{2,IGIMF}$ is taking higher values i.e.
steeper slopes for z = 0 - 2. But gradually due to the rise of
temperature of the medium with increasing z, formation of massive
stars predominates even for a comparatively lower but still
moderate SFR and hence for moderate $M_{ecl,max}$. This favors
formation of massive stars which makes $\alpha_{2,IGIMF}$ lower
for $z \sim 2 - 4$. The effect becomes reduced due to rapid fall
of SFR at very high z (viz. $z \sim
4-6$, Figs. 1 - 4) increasing $\alpha_{2,IGIMF}$ indices again.\\

\noindent (ii) As $\beta$ increases, changes in the rising and
falling of $\alpha_{2,IGIMF}$  become faster. The effect is very
pronounced for $\beta \sim 2 - 2.4$. This is because, as $\beta$
increases, the number of low mass clusters become higher compared
to the number of high mass clusters. So the above mentioned effect
becomes accentuated due to steepening of the mass
function of embedded clusters (Fig. 6).
{\bf The statistical test has shown rejection of the null hypothesis.  }\\

\noindent (iii) $\alpha_{2,IGIMF}$ becomes flatter as
$M_{ecl,min}$ increases when $\beta$ is low. This is because when
$\beta$ is low, number of low mass clusters  decreases and as a
result massive star formation is favored compared to low mass
stars. This results in the flattening of the slopes,
$\alpha_{2,IGIMF}$ in high mass regime. For higher values of
$\beta \ (>$ 2), number of low mass  clusters increases which
disfavors formation of massive stars and
$\alpha_{2,IGIMF}$ becomes steeper as a result.\\

\noindent (iv) $\alpha_{2,IGIMF}$  is always flatter than the IMF
slopes. This might be the joint effect of various star formation
rates as well as increasing temperature of the environment with
increasing redshifts. Up to $z \sim 2 $ the temperature of the
ambient medium is lower compared to the higher redshift zone.
Hence Jeans masses are lower. But at the same time SFR is
increasing to its maximum increasing $M_{ecl,max}$ which favors
over all formation of massive stars compared to IMF. On the other
hand for $z
> 2 $, Jeans masses are higher and SFR gradually decrease lowering
values of $M_{ecl,max}$, hence increasing low mass stars. Somehow
the joint effect of these two phenomena is responsible for a
resulting flatness of $\alpha_{2,IGIMF}$. This is consistent with
some observational results (Alonso et al. 2004; Finoguenov et al.
2003; Lowenstein 2006; Nayakshin \& Sunyaev 2005) which indicate
IGIMF to be top heavy (i.e. massive stars form in large numbers
compared to less massive stars) when SFR $\ge$ 100
$M_{\odot}yr^{-1}$. The above trend is also in good agreement for
Galactic and M31 bulge (Ballero et al. 2007) as well as Wilkin et
al. (2011) for present day mass density from cosmological star
formation history. The trend for decreasing slope with increasing
SFR has also been found by Gunawardhana et al. (2011) for a sample
of 40000 galaxies. For $z>2$, though the SFR decreases and
formation of massive clouds are not favored, but still massive
stars are produced in some optimum zone due to the increase in
temperature of the medium so that $m_{c}^{'}$ is shifted towards
higher mass (i.e. a top heavy mass spectrum with
steeper slope ) (viz. Figs.9 - 11).\\

\noindent (v) The characteristic mass $m_{c}$ of stellar initial
mass function differs from characteristic mass, $m_{c^{'}}$, of
integrated galactic mass function. Generally $m_{c} \geq m_{c^{'}}$.\\

\noindent (vi) the above mentioned effects are similar for various
measures of SFR though there are small variations. The measure
SFR1 is the first quartile i.e. 25 \% of the galaxies have $SFR
\le SFR1$ and 75\% of the galaxies have $SFR > SFR1$ i.e. we can
say SFR1 is representative one for low SFR which is the
characteristics of dwarf galaxies. On the other hand SFR3 is
representative one of high SFR which characterizes giant galaxies.
In this regard SFR2 is the measure of average SFR of galaxies. Now
in dwarf galaxies due to its low SFR, formation of massive stars
are not favorable in large numbers. Thus it is most likely that
$\alpha_{2,IGIMF}$ for SFR1 is rather steeper than that of
$\alpha_{2,IGIMF}$ for SFR2 followed by $\alpha_{2,IGIMF}$ for
SFR3. It is clear from Tables 2 - 4 that $\alpha_{2,IGIMF}$ for
SFR1 $>$ $\alpha_{2,IGIMF}$ for SFR2 in 72 \% - 78 \% cases for
$\beta $ = 2 - 2.6 and $\alpha_{2,IGIMF}$ for SFR2 $>$
$\alpha_{2,IGIMF}$ for SFR3  in 50 \% - 60 \% cases for $\beta$ =
2 - 2.6. SFR$^*$ is the point of the SFR function where the
function levels off from exponential to a shallower power law i.e.
from this point the SFR does not vary much to the left i.e. for
low SFR. So SFR$^*$ is sort of representative value of low SFR
i.e. of less massive galaxies. $\alpha_{2,IGIMF}$ for SFR$^*$ $>$
$\alpha_{2,IGIMF}$ for SFR2 in 67\% - 50 \% cases for $\beta =$ 2
- 2.6. Therefore SFR1/ SFR$^*$, SFR2, SFR3  might be
representative star formation histories for dwarf, intermediate
and giant galaxies {\bf and hence the selection of the masses of
the galaxies is appropriate.}

\section{Conclusion}

\noindent  In the present work for the first time the nature of
observed star formation rate has been investigated (viz. SFR1,
SFR2, SFR3 and SFR$^*$) as a function of redshift instead of {\bf
using SFR as a parameter by some authors (Weidner \& Kroupa 2004)}
for various types of galaxies (viz. dwarf, intermediate and giant)
together with a top heavy stellar IMF {\bf increasing  with
redshift (viz. equation (6))}. This helps to study the cosmic star
formation history in galaxies under the combining effect of both
varying IMF and SFR. A Monte Carlo simulation method is used for
its simplicity for computation to find the resulting IGIMF. It is
found that up to a redshift of z $\sim$ 2, the galactic mass
function becomes steeper compared to a flatter one for $z> 2$
followed again by a steeper one around $z \sim 6$. This is due to
the joint effect of the distribution of SFR as a function of z and
temperature of the ambient medium. The galactic mass function is
affected by the embedded cluster mass-spectrum . The effect is
faster for a steeper one. It is also influenced by the minimum
mass of the parent cluster e.g. $\alpha_{2,IGIMF}$ becomes flatter
as $M_{ecl,min}$ increases when $\beta$ is low.

\clearpage

\section{Acknowledgements}

\noindent  The authors are very much thankful for the suggestions
of the referee to improve the quality of the manuscript to a great
extent. One of the the authors (Tanuka Chattopadhyay) wishes to
thank Department of Science and Technology (DST), India for
awarding her a major research project for the work. The authors
are also grateful to Soumita Modak for her help.

\clearpage

\begin{figure}
\centering
\includegraphics[width=0.7\textwidth]{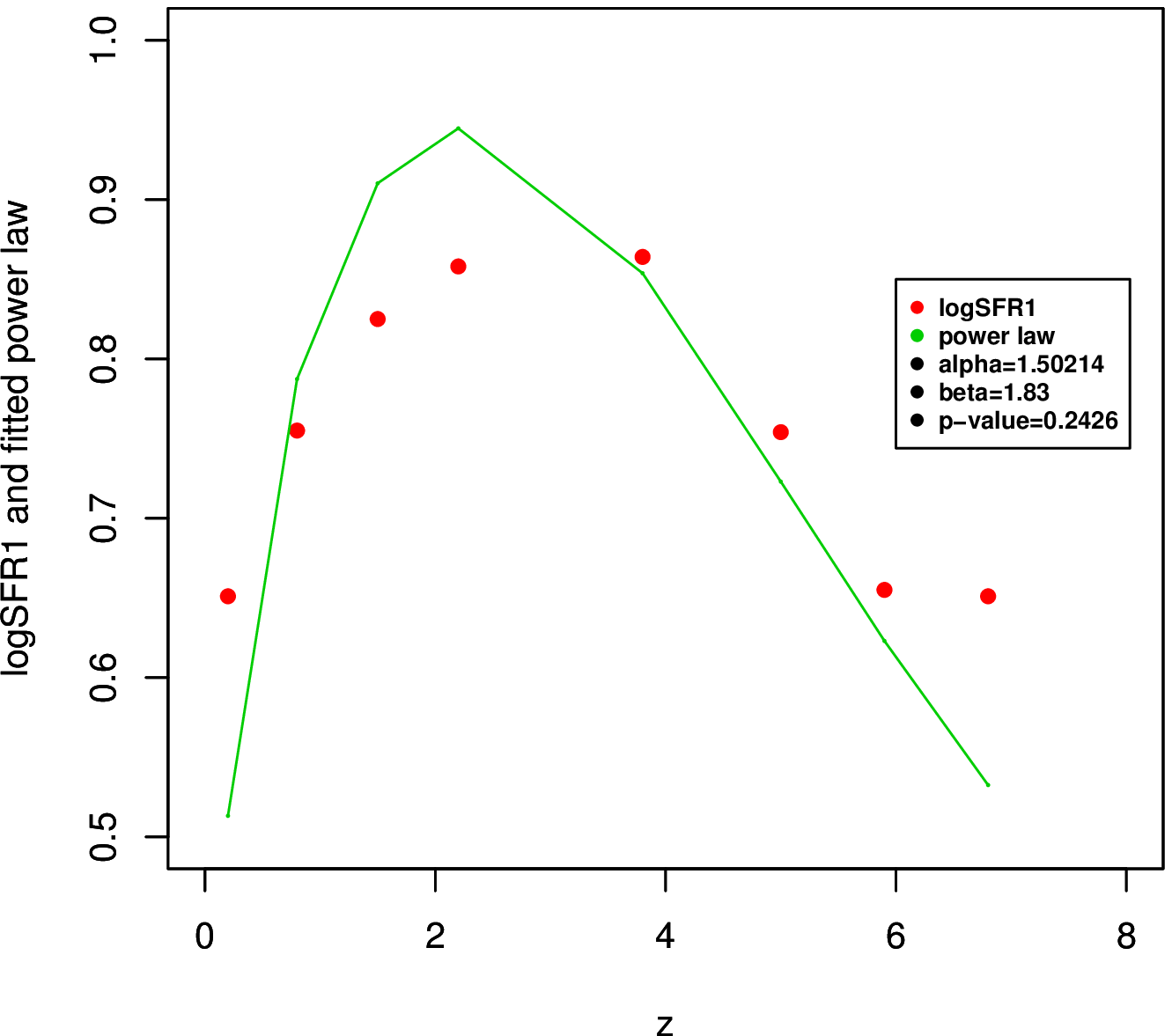}
\caption{Tapered power law fit to the observed SFR1 as
$log10(SFR1) = 10.0333*(z^{-1.5021})*(1-exp(-z/3.8)^{1.83})$,
p-value = 0.2426) } \label{Figure 1}
\end{figure}
\clearpage

\begin{figure}
\centering
\includegraphics[width=0.7\textwidth]{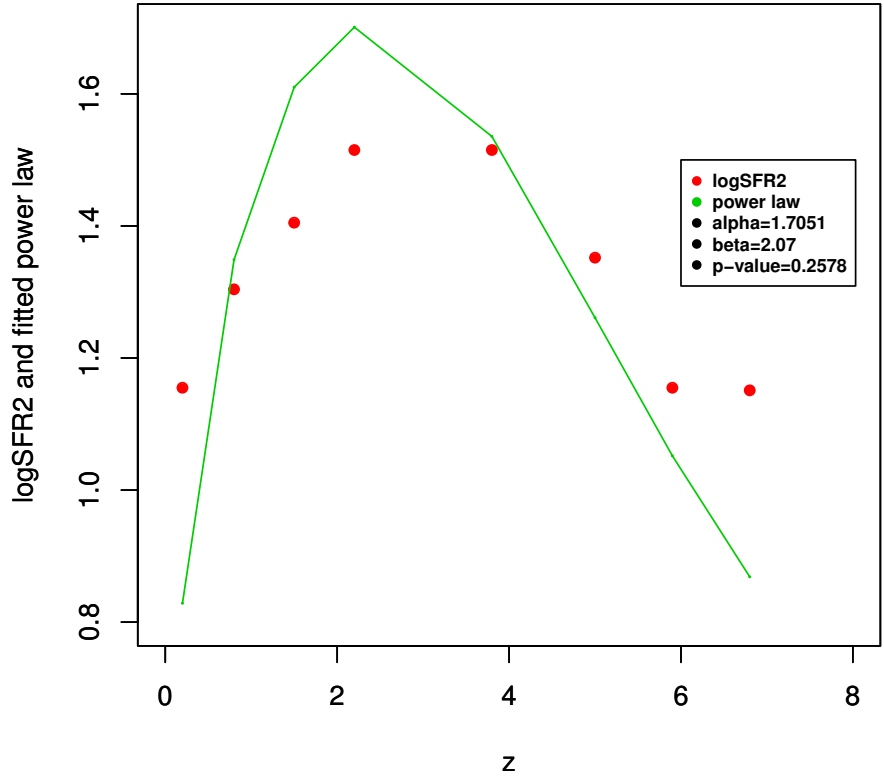}
\caption{Tapered power law fit to the observed SFR2 as
$log10(SFR2) = 23.6636*(z^{-1.7051)}*(1-exp(-(z/3.8)^{2.07})$,
p-value = 0.2578) } \label{Figure 2}
\end{figure}
\clearpage

\begin{figure}
\centering
\includegraphics[width=0.7\textwidth]{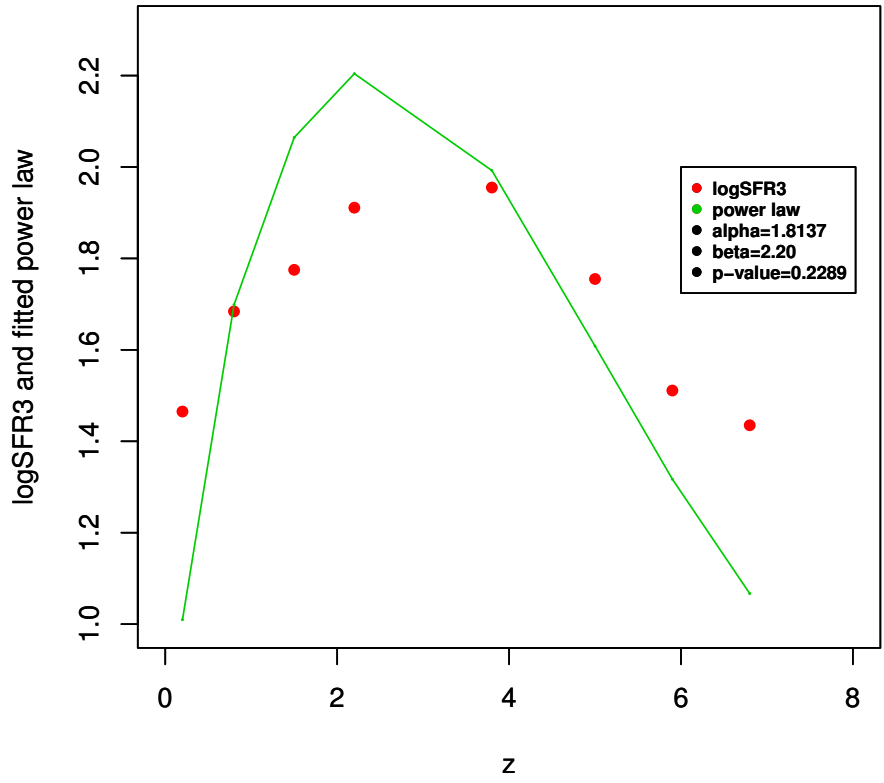}
\caption{Tapered power law fit to the observed SFR3 as
$log10(SFR3) = 35.4905*(z^{-1.8137})*(1-exp(-(z/3.8)^{2.20})$,
p-value = 0.2503) } \label{Figure 3}
\end{figure}
\clearpage

\begin{figure}
\centering
\includegraphics[width=0.7\textwidth]{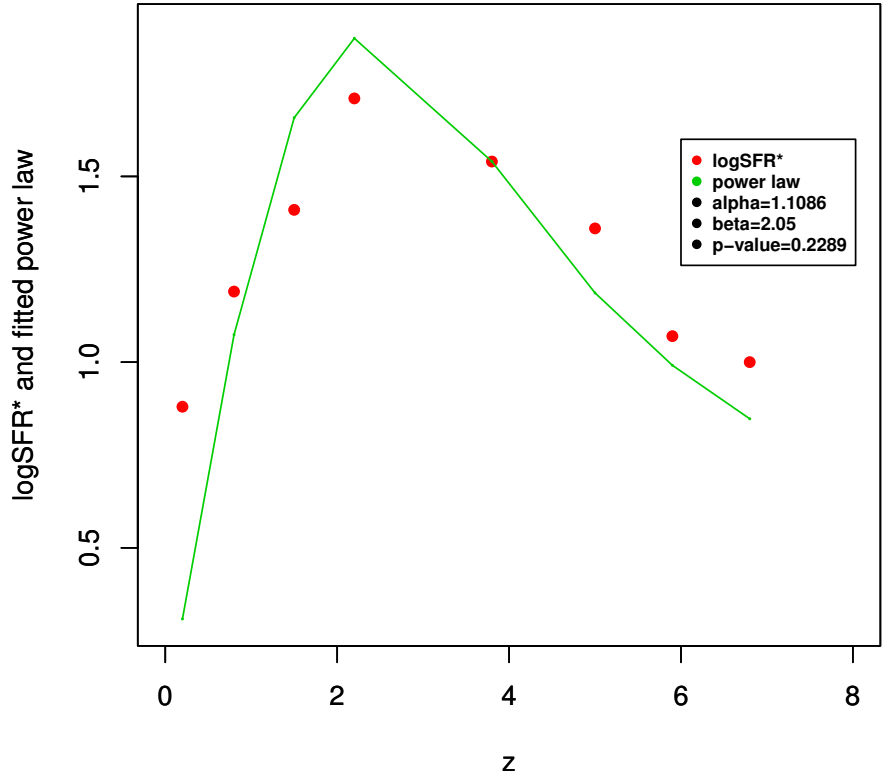}
\caption{Tapered power law fit to the observed SFR$^*$ as
$log10(SFR^*) = 7.097*(z^{-1.1086})*(1-exp(-(z/2.2)^{2.05}),$
p-value = 0.2289). } \label{Figure 4}
\end{figure}

\clearpage

\begin{figure}
\centering
\includegraphics[width=0.7\textwidth]{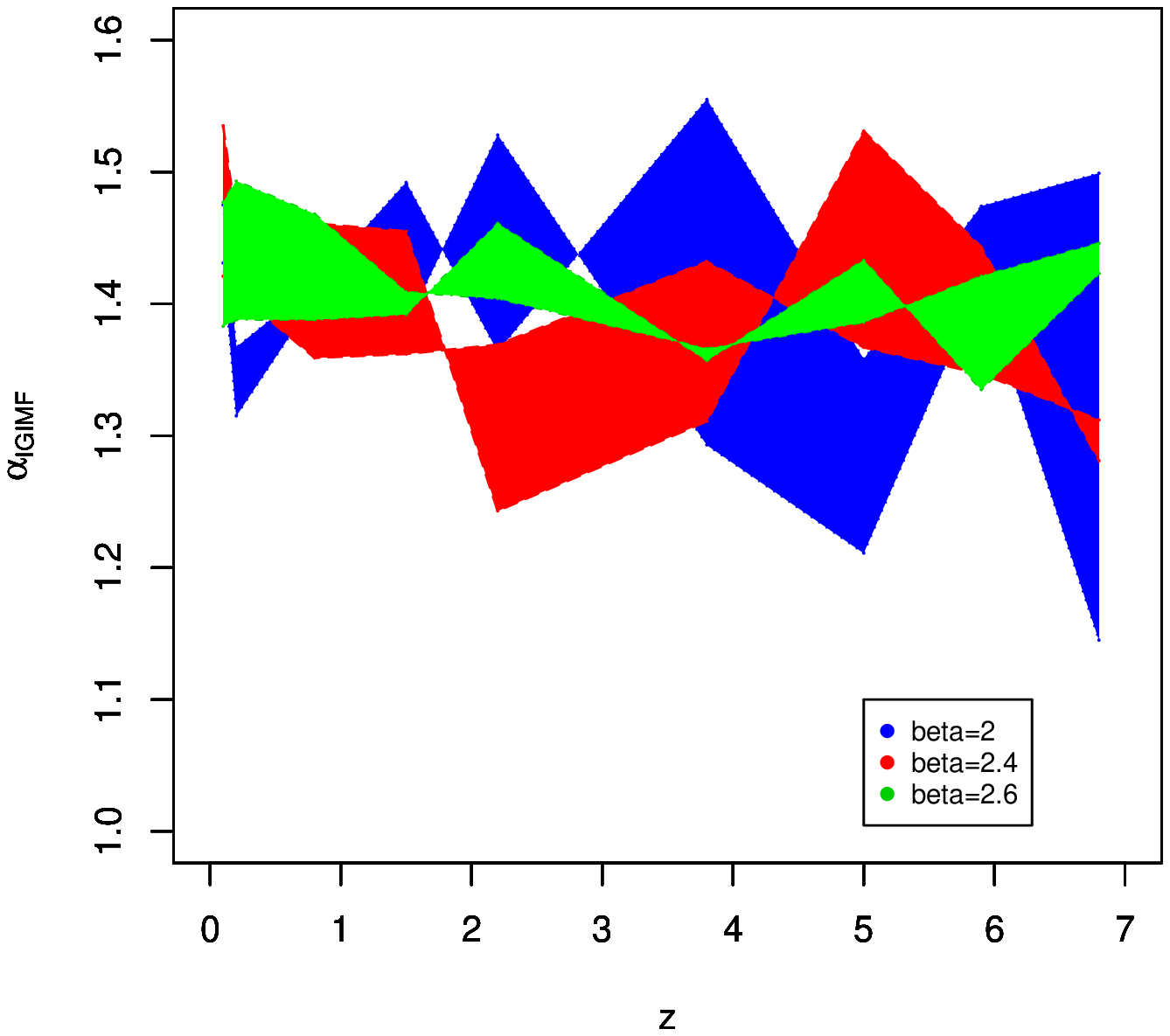}
\caption{$\alpha_{2,IGIMF}$ as a function of z for all values of
parameters given in Tables 2-3 for SFR1 and $\beta$ = 2.6 (green
region), $\beta$ = 2.4 (red region), $\beta $= 2 (blue
region).}\label{Figure 5}
\end{figure}

\clearpage

\begin{figure}
\centering
\includegraphics[width=0.7\textwidth]{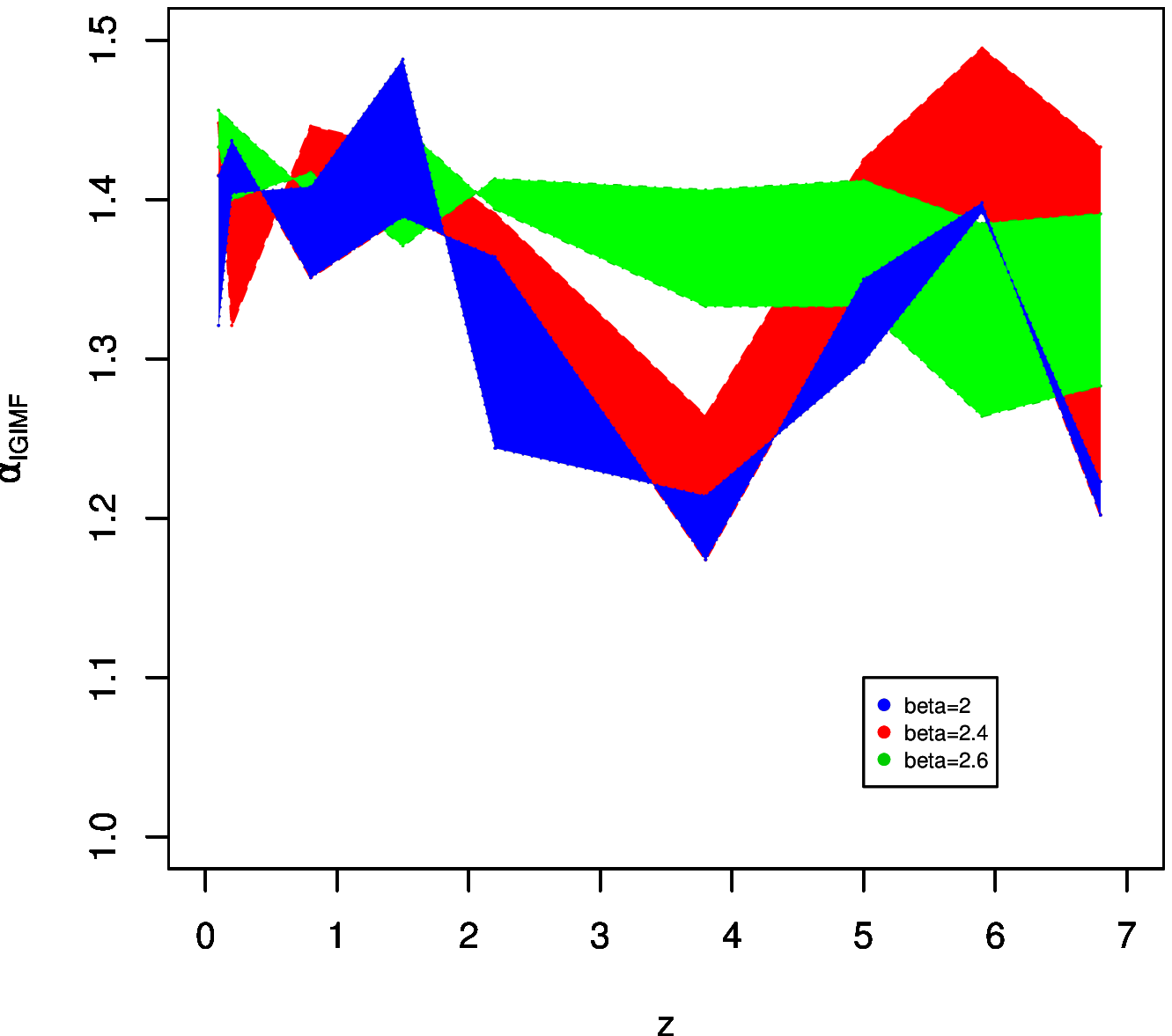}
\caption{$\alpha_{2,IGIMF}$ as a function of z for all values of
parameters given in Tables 4-5 for SFR2 and $\beta$ = 2.6 (green
region), $\beta$ = 2.4 (red region), $\beta$ = 2 (blue
region).}\label{Figure 6}
\end{figure}

\clearpage

\begin{figure}
\centering
\includegraphics[width=0.7\textwidth]{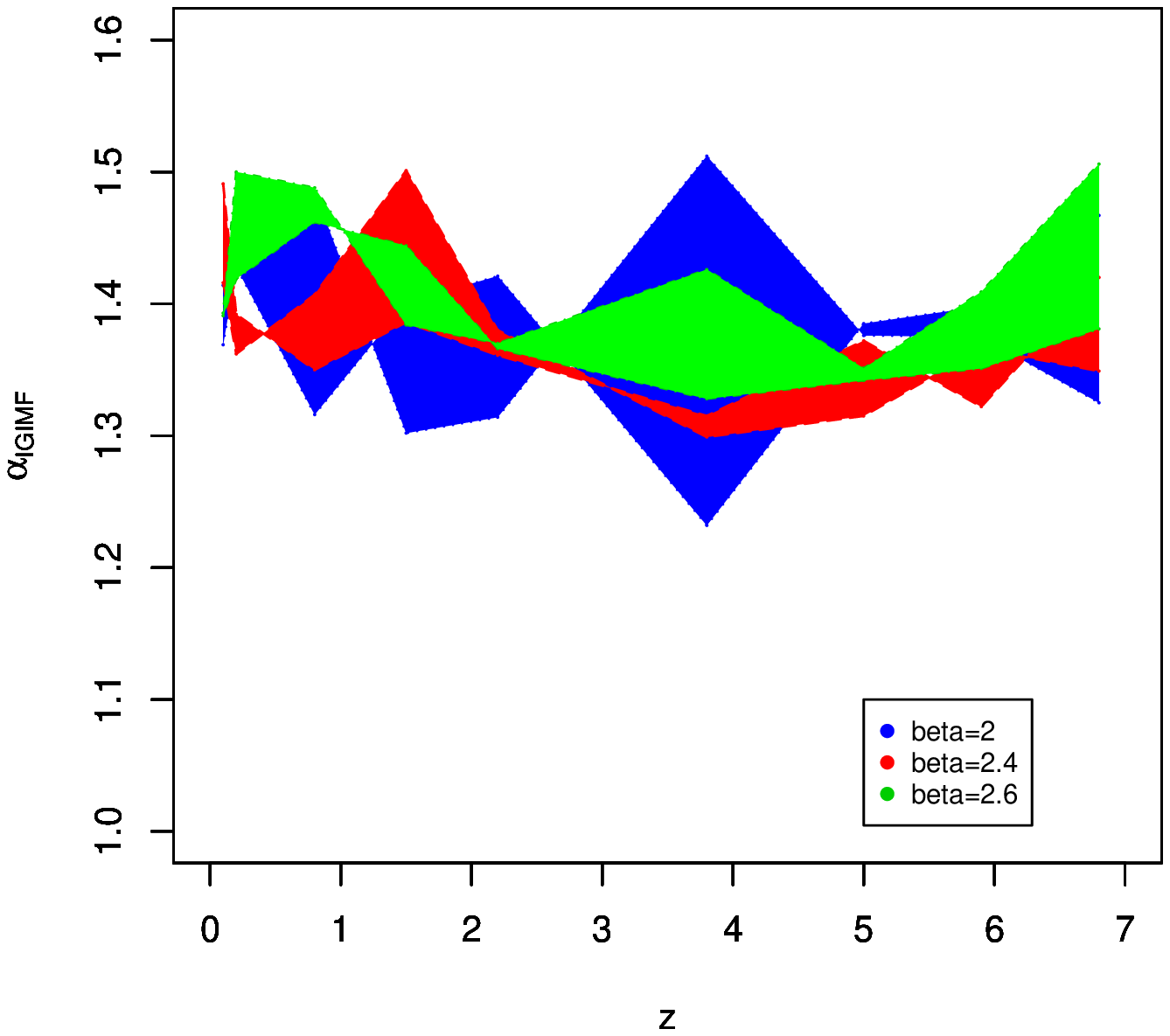}
\caption{$\alpha_{2,IGIMF}$ as a function of z for all values of
parameters given in Tables 6-7 for SFR3 and $\beta$ = 2.6 (green
region), $\beta$ = 2.4 (red region), $\beta$ = 2 (blue
region).}\label{Figure 7}
\end{figure}

\clearpage

\begin{figure}
\centering
\includegraphics[width=0.7\textwidth]{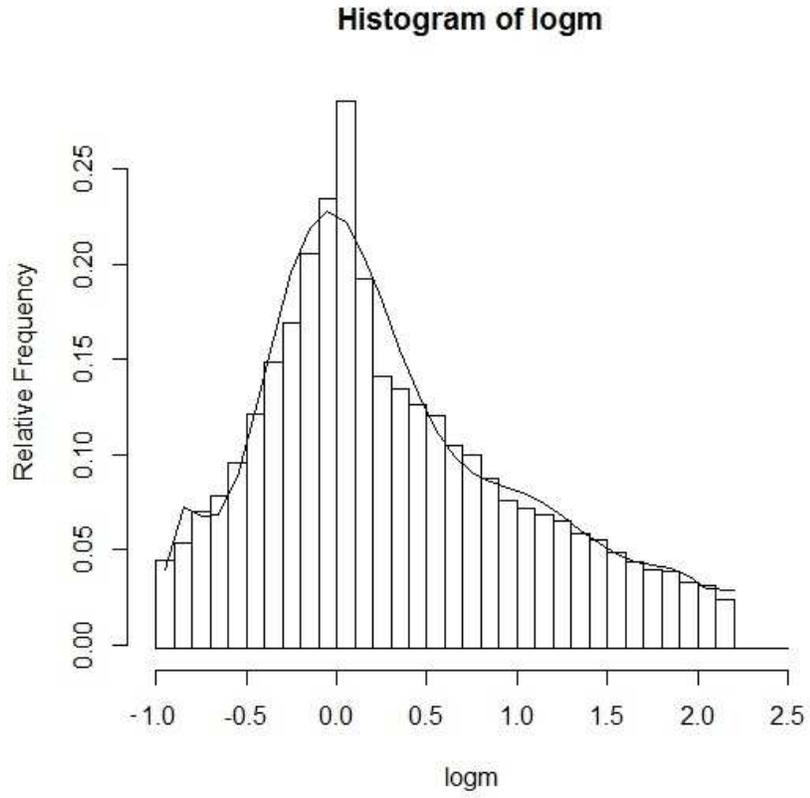}
\caption{One representative histogram with fitted curve for  the
simulated points of the IGIMF ($\xi (m) =\frac{dN}{dlogm}$) for
redshift z = 1.5, $\beta$ = 2, $Mecl, min=500 M_{\odot}$ and
SFR2.} \label{Figure 8}
\end{figure}
\clearpage

\begin{figure}
\centering
\includegraphics[width=0.7\textwidth]{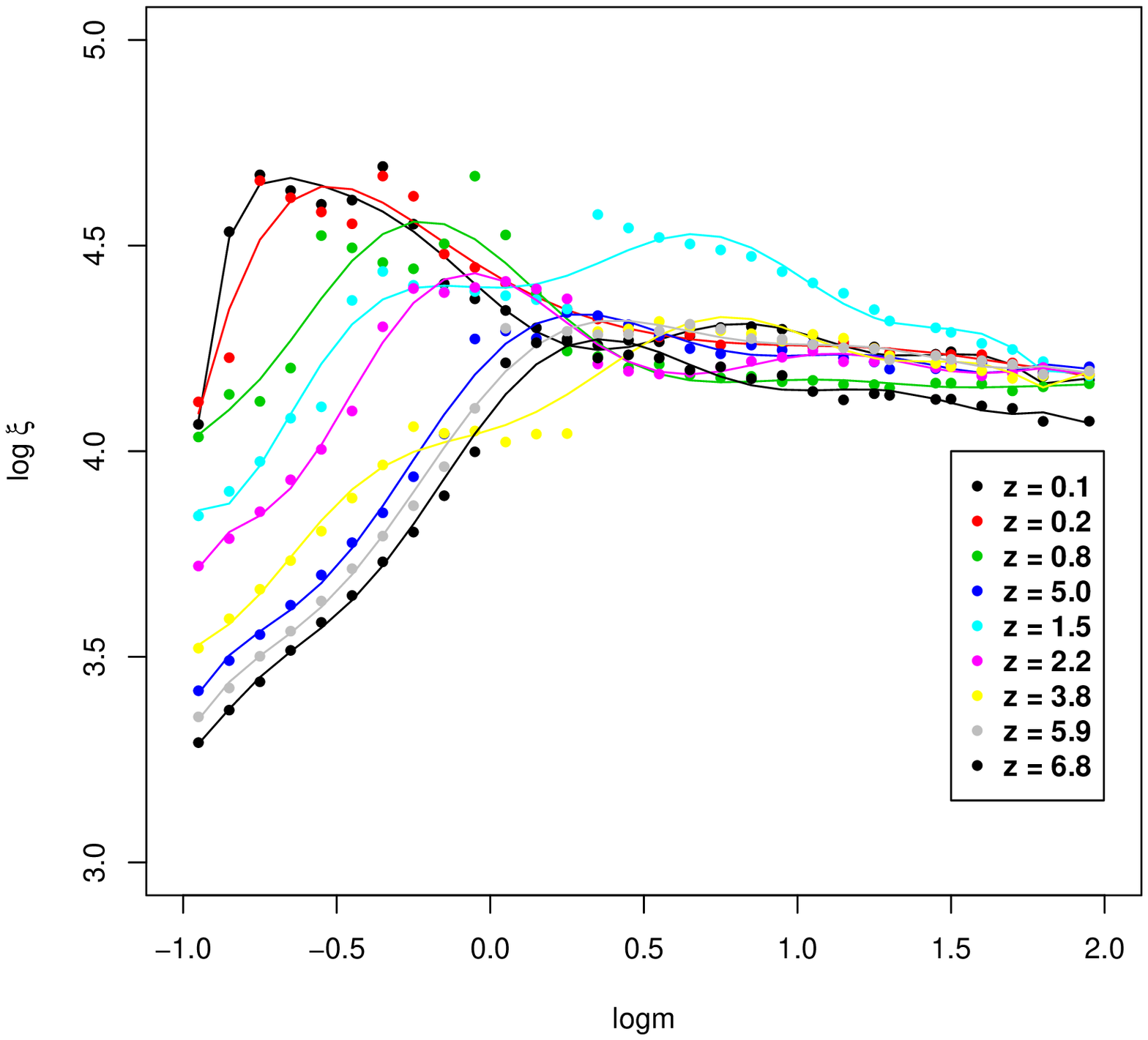}
\caption{Fitted curves for  the simulated points of the IGIMF
($\xi (m) =\frac{dN}{dlogm}$) vs logm for various redshifts at
$\beta$ = 2 for $Mecl, min=500 M_{\odot}$ and SFR1.} \label{Figure
8}
\end{figure}
\clearpage

\begin{figure}
\centering
\includegraphics[width=0.7\textwidth]{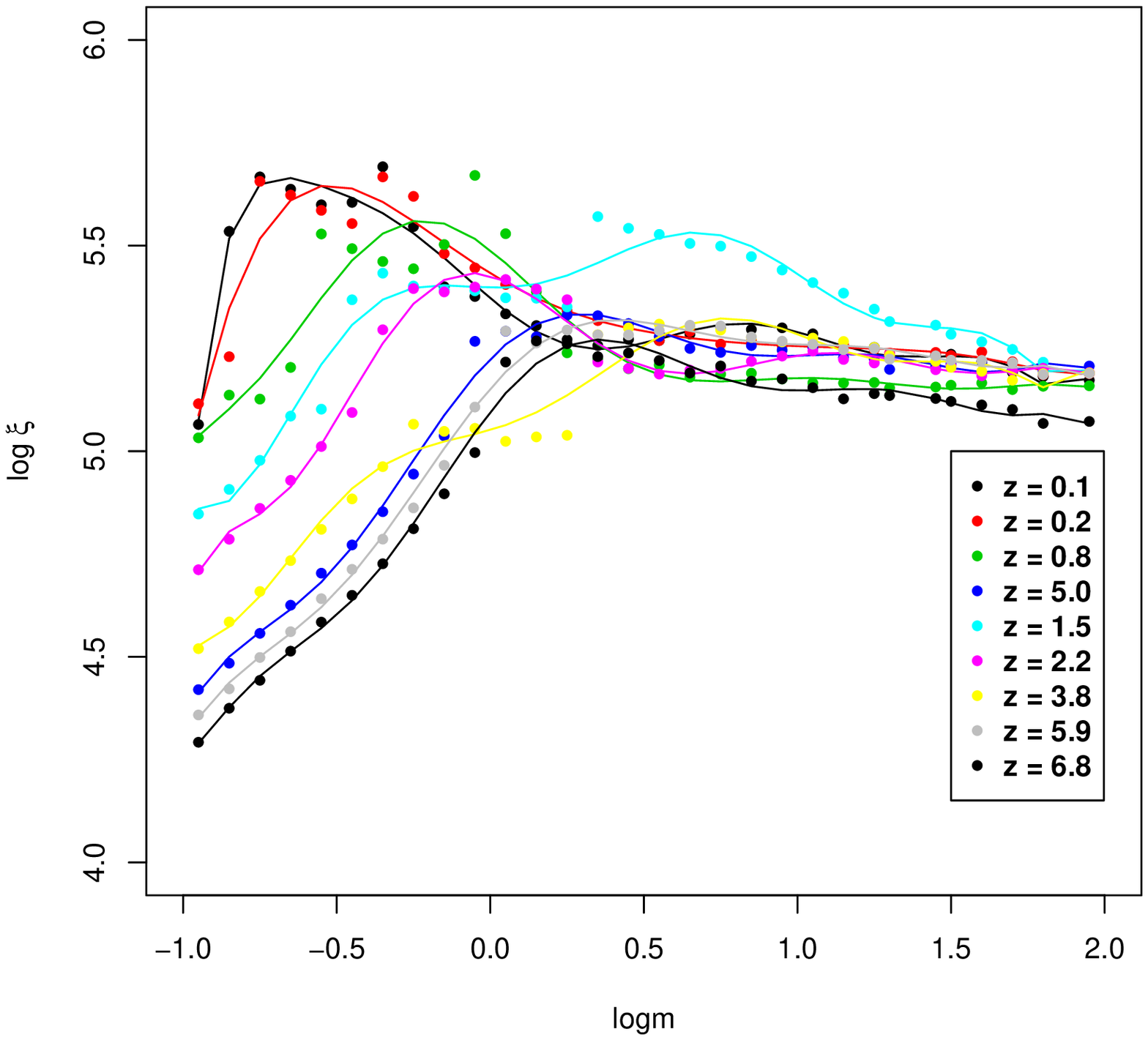}
\caption{Fitted curves for  the simulated points of the IGIMF
($\xi(m)=\frac{dN}{dlogm}$) vs logm for various redshifts at
$\beta$ = 2 for $Mecl ,min=500 M_{\odot}$ and SFR2.} \label{Figure
9}
\end{figure}
\clearpage

\begin{figure}
\centering
\includegraphics[width=0.7\textwidth]{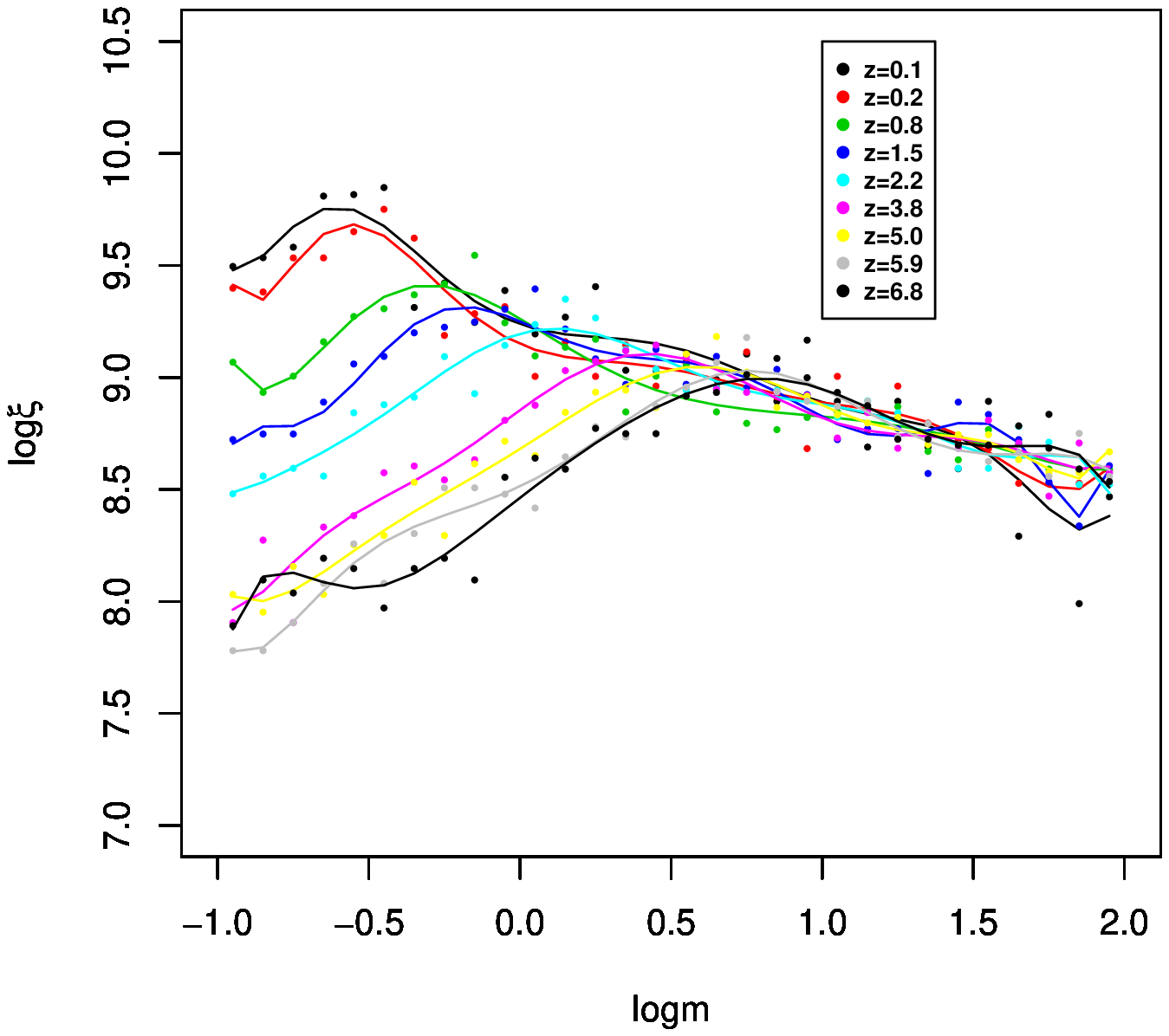}
\caption{Fitted curves for  the simulated points of the IGIMF
($\xi(m)=\frac{dN}{dlogm}$) vs logm for various redshifts at
$\beta$ = 2 for $Mecl, min=500 M_{\odot}$ and SFR3.} \label{Figure
10}
\end{figure}
\clearpage

\clearpage
\begin{table}
\centering \caption{Initial values of the parameters}
\begin{tabular}{@{}cc}
\hline \hline
Parameter & value   \\
\hline
$\alpha_{1,IMF}(z=0)$ & 1.25 \\
$\alpha_{2,IMF}(z=0)$ &2.35 \\
Galaxy masses & $5\times10^{9}M_{\bigodot}$, $5\times10^{10}M_{\bigodot}$, $5\times10^{11}M_{\bigodot}$\\
$m_{min}$ &$0.1M_{\bigodot}$\\
$m_{max}$ &$150M_{\bigodot}$\\
$m_{c,IMF}(z=0)$ & $0.3M_{\bigodot}$\\
$M_{ecl,min}$ &  500,1000$(M_{\bigodot})$\\
$\beta$ & 2.0,2.4,2.6\\
z & 0.1,0.2,0.8,1.5,2.2,3.8,5.0,5.9,6.8\\
efficiency & 30$\%$\\ \hline \hline
\end{tabular}
\end{table}
\clearpage

\clearpage
\begin{table}
\begin{tiny}

\caption{IGIMF and IMF slopes with varying z and $Mecl_{min}$ at
$\beta$=2,2.4 for SFR1:}
 \label{tab:final2}
\begin{tabular*}{7.055in}{ccccccccccccc}
\hline
\multicolumn{13}{c}{\textbf{$\beta=2.0$}}\\
 \hline
 &\multicolumn{6}{c}{$z=0.1$}&\multicolumn{6}{c}{$z=0.2$}\\
 \hline
$Mecl_{min}$&$\alpha_{1,IMF}$&$\alpha_{2,IMF}$&$m_{c}$&$\alpha_{1,IGIMF}$&$\alpha_{2,IGIMF}$&$m_{c^{'}}$
&$\alpha_{1,IMF}$&$\alpha_{2,IMF}$&$m_{c}$&$\alpha_{1,IGIMF}$&$\alpha_{2,IGIMF}$&$m_{c^{'}}$\\
\hline
500 & 1.25 & 2.35 & 0.381 & -0.05 & 1.475 & 0.224 & 1.25 & 2.35 & 0.434 & 0.161 & 1.367 & 0.282 \\
1000 & 1.25 & 2.35 & 0.381 & 0.072 & 1.431 & 0.178 & 1.25 & 2.35 & 0.434 & -0.303 & 1.315 & 0.282 \\
 \hline
 &\multicolumn{6}{c}{$z=0.8$}&\multicolumn{6}{c}{$z=1.5$}\\
 \hline
$Mecl_{min}$&$\alpha_{1,IMF}$&$\alpha_{2,IMF}$&$m_{c}$&$\alpha_{1,IGIMF}$&$\alpha_{2,IGIMF}$&$m_{c^{'}}$
&$\alpha_{1,IMF}$&$\alpha_{2,IMF}$&$m_{c}$&$\alpha_{1,IGIMF}$&$\alpha_{2,IGIMF}$&$m_{c^{'}}$\\
\hline
500 & 1.25 & 2.35 & 0.797 & 0.235 & 1.415 & 0.562 & 1.25 & 2.35 & 1.304 & 0.178 & 1.492 & 0.891 \\
1000 & 1.25 & 2.35 & 0.797 & 0.305 & 1.404 & 0.355 & 1.25 & 2.35 & 1.304 & 0.095 & 1.384 & 0.891 \\
\hline
 &\multicolumn{6}{c}{$z=2.2$}&\multicolumn{6}{c}{$z=3.8$}\\
 \hline
$Mecl_{min}$&$\alpha_{1,IMF}$&$\alpha_{2,IMF}$&$m_{c}$&$\alpha_{1,IGIMF}$&$\alpha_{2,IGIMF}$&$m_{c^{'}}$
&$\alpha_{1,IMF}$&$\alpha_{2,IMF}$&$m_{c}$&$\alpha_{1,IGIMF}$&$\alpha_{2,IGIMF}$&$m_{c^{'}}$\\
\hline
500 & 1.25 & 2.35 & 1.889 & 0.315 & 1.555 & 1.412 & 1.25 & 2.35 & 3.471 & 0.168 & 1.358 & 2.239 \\
1000 & 1.25 & 2.35 & 1.889 & 0.224 & 1.528 & 1.122 & 1.25 & 2.35 & 3.471 & 0.505 & 1.293 & 2.239 \\
\hline
 &\multicolumn{6}{c}{$z=5.0$}&\multicolumn{6}{c}{$z=5.9$}\\
 \hline
$Mecl_{min}$&$\alpha_{1,IMF}$&$\alpha_{2,IMF}$&$m_{c}$&$\alpha_{1,IGIMF}$&$\alpha_{2,IGIMF}$&$m_{c^{'}}$
&$\alpha_{1,IMF}$&$\alpha_{2,IMF}$&$m_{c}$&$\alpha_{1,IGIMF}$&$\alpha_{2,IGIMF}$&$m_{c^{'}}$\\
\hline
500 & 1.25 & 2.35 & 4.849 & 0.237 & 1.474 & 4.467 & 1.25 & 2.35 & 5.981 & 0.522 & 1.499 & 11.22 \\
1000 & 1.25 & 2.35 & 4.849 & 0.354 & 1.211 & 4.467 & 1.25 & 2.35 & 5.981 & 0.261 & 1.416 & 4.467 \\
 \hline
 &\multicolumn{6}{c}{$z=6.8$}\\
 \hline
$Mecl_{min}$&$\alpha_{1,IMF}$&$\alpha_{2,IMF}$&$m_{c}$&$\alpha_{1,IGIMF}$&$\alpha_{2,IGIMF}$&$m_{c^{'}}$
\\
\hline
500 & 1.25 & 2.35 & 7.189 & 0.348 & 1.409 & 11.22 &  &  &  &  &  &  \\
1000 & 1.25 & 2.35 & 7.189 & 0.312 & 1.145 & 7.079 &  &  &  &  &  &  \\
\hline
\multicolumn{13}{c}{\textbf{$\beta=2.4$}}\\
 \hline
 &\multicolumn{6}{c}{$z=0.1$}&\multicolumn{6}{c}{$z=0.2$}\\
 \hline
$Mecl_{min}$&$\alpha_{1,IMF}$&$\alpha_{2,IMF}$&$m_{c}$&$\alpha_{1,IGIMF}$&$\alpha_{2,IGIMF}$&$m_{c^{'}}$
&$\alpha_{1,IMF}$&$\alpha_{2,IMF}$&$m_{c}$&$\alpha_{1,IGIMF}$&$\alpha_{2,IGIMF}$&$m_{c^{'}}$\\
\hline
500 & 1.25 & 2.35 & 0.381 & 0.654 & 1.421 & 0.224 & 1.25 & 2.35 & 0.434 & 0.254 & 1.409 & 0.282 \\
1000 & 1.25 & 2.35 & 0.381 & 0.645 & 1.535 & 0.224 & 1.25 & 2.35 & 0.434 & 0.335 & 1.462 & 0.282 \\
\hline
 &\multicolumn{6}{c}{$z=0.8$}&\multicolumn{6}{c}{$z=1.5$}\\
 \hline
$Mecl_{min}$&$\alpha_{1,IMF}$&$\alpha_{2,IMF}$&$m_{c}$&$\alpha_{1,IGIMF}$&$\alpha_{2,IGIMF}$&$m_{c^{'}}$
&$\alpha_{1,IMF}$&$\alpha_{2,IMF}$&$m_{c}$&$\alpha_{1,IGIMF}$&$\alpha_{2,IGIMF}$&$m_{c^{'}}$\\
\hline
500 & 1.25 & 2.35 & 0.797 & 0.205 & 1.359 & 0.562 & 1.25 & 2.35 & 1.304 & 0.163 & 1.362 & 1.122 \\
1000 & 1.25 & 2.35 & 0.797 & 0.226 & 1.461 & 0.562 & 1.25 & 2.35 & 1.304 & 0.204 & 1.455 & 0.891 \\
\hline
 &\multicolumn{6}{c}{$z=2.2$}&\multicolumn{6}{c}{$z=3.8$}\\
 \hline
$Mecl_{min}$&$\alpha_{1,IMF}$&$\alpha_{2,IMF}$&$m_{c}$&$\alpha_{1,IGIMF}$&$\alpha_{2,IGIMF}$&$m_{c^{'}}$
&$\alpha_{1,IMF}$&$\alpha_{2,IMF}$&$m_{c}$&$\alpha_{1,IGIMF}$&$\alpha_{2,IGIMF}$&$m_{c^{'}}$\\
\hline
500 & 1.25 & 2.35 & 1.889 & 0.255 & 1.369 & 1.412 & 1.25 & 2.35 & 3.471 & 0.344 & 1.432 & 2.818 \\
1000 & 1.25 & 2.35 & 1.889 & 0.215 & 1.243 & 1.412 & 1.25 & 2.35 & 3.471 & 0.294 & 1.311 & 4.467 \\
 \hline
 &\multicolumn{6}{c}{$z=5.0$}&\multicolumn{6}{c}{$z=5.9$}\\
 \hline
$Mecl_{min}$&$\alpha_{1,IMF}$&$\alpha_{2,IMF}$&$m_{c}$&$\alpha_{1,IGIMF}$&$\alpha_{2,IGIMF}$&$m_{c^{'}}$
&$\alpha_{1,IMF}$&$\alpha_{2,IMF}$&$m_{c}$&$\alpha_{1,IGIMF}$&$\alpha_{2,IGIMF}$&$m_{c^{'}}$\\
\hline
500 & 1.25 & 2.35 & 4.849 & 0.185 & 1.367 & 3.548 & 1.25 & 2.35 & 5.981 & 0.183 & 1.347 & 4.467 \\
1000 & 1.25 & 2.35 & 4.849 & 0.239 & 1.531 & 2.818 & 1.25 & 2.35 & 5.981 & 0.141 & 1.443 & 4.467 \\
\hline
 &\multicolumn{6}{c}{$z=6.8$}\\
 \hline
$Mecl_{min}$&$\alpha_{1,IMF}$&$\alpha_{2,IMF}$&$m_{c}$&$\alpha_{1,IGIMF}$&$\alpha_{2,IGIMF}$&$m_{c^{'}}$
\\
\hline
500 & 1.25 & 2.35 & 7.189 & 0.291 & 1.312 & 7.079 &  &  &  &  &  &  \\
1000 & 1.25 & 2.35 & 7.189 & 0.249 & 1.281 & 7.079 &  &  &  &  &  &  \\
\hline
\end{tabular*}
\end{tiny}
\end{table}
\clearpage

\begin{table}
\begin{tiny}

\caption{IGIMF and IMF slopes with varying z and $Mecl_{min}$ at
$\beta$=2.6 for SFR1:}
 \label{tab:final2}
\begin{tabular*}{7.055in}{ccccccccccccc}
\hline
\multicolumn{13}{c}{\textbf{$\beta=2.6$}}\\
 \hline
 &\multicolumn{6}{c}{$z=0.1$}&\multicolumn{6}{c}{$z=0.2$}\\
 \hline
$Mecl_{min}$&$\alpha_{1,IMF}$&$\alpha_{2,IMF}$&$m_{c}$&$\alpha_{1,IGIMF}$&$\alpha_{2,IGIMF}$&$m_{c^{'}}$
&$\alpha_{1,IMF}$&$\alpha_{2,IMF}$&$m_{c}$&$\alpha_{1,IGIMF}$&$\alpha_{2,IGIMF}$&$m_{c^{'}}$\\
\hline
500 & 1.25 & 2.35 & 0.381 & 0.072 & 1.383 & 0.224 & 1.25 & 2.35 & 0.434 & 0.495 & 1.388 & 0.282 \\
1000 & 1.25 & 2.35 & 0.381 & 0.01 & 1.477 & 0.224 & 1.25 & 2.35 & 0.434 & 0.579 & 1.493 & 0.282 \\
\hline
 &\multicolumn{6}{c}{$z=0.8$}&\multicolumn{6}{c}{$z=1.5$}\\
 \hline
$Mecl_{min}$&$\alpha_{1,IMF}$&$\alpha_{2,IMF}$&$m_{c}$&$\alpha_{1,IGIMF}$&$\alpha_{2,IGIMF}$&$m_{c^{'}}$
&$\alpha_{1,IMF}$&$\alpha_{2,IMF}$&$m_{c}$&$\alpha_{1,IGIMF}$&$\alpha_{2,IGIMF}$&$m_{c^{'}}$\\
\hline
500 & 1.25 & 2.35 & 0.797 & 0.281 & 1.388 & 0.562 & 1.25 & 2.35 & 1.304 & 0.286 & 1.392 & 0.891 \\
1000 & 1.25 & 2.35 & 0.797 & 0.083 & 1.468 & 0.708 & 1.25 & 2.35 & 1.304 & 0.424 & 1.409 & 0.891 \\
 \hline
 &\multicolumn{6}{c}{$z=2.2$}&\multicolumn{6}{c}{$z=3.8$}\\
 \hline
$Mecl_{min}$&$\alpha_{1,IMF}$&$\alpha_{2,IMF}$&$m_{c}$&$\alpha_{1,IGIMF}$&$\alpha_{2,IGIMF}$&$m_{c^{'}}$
&$\alpha_{1,IMF}$&$\alpha_{2,IMF}$&$m_{c}$&$\alpha_{1,IGIMF}$&$\alpha_{2,IGIMF}$&$m_{c^{'}}$\\
\hline
500 & 1.25 & 2.35 & 1.889 & 0.242 & 1.461 & 1.412 & 1.25 & 2.35 & 3.471 & 0.319 & 1.357 & 2.239 \\
1000 & 1.25 & 2.35 & 1.889 & 0.218 & 1.404 & 1.778 & 1.25 & 2.35 & 3.471 & 0.256 & 1.366 & 2.818 \\
\hline
 &\multicolumn{6}{c}{$z=5.0$}&\multicolumn{6}{c}{$z=5.9$}\\
 \hline
$Mecl_{min}$&$\alpha_{1,IMF}$&$\alpha_{2,IMF}$&$m_{c}$&$\alpha_{1,IGIMF}$&$\alpha_{2,IGIMF}$&$m_{c^{'}}$
&$\alpha_{1,IMF}$&$\alpha_{2,IMF}$&$m_{c}$&$\alpha_{1,IGIMF}$&$\alpha_{2,IGIMF}$&$m_{c^{'}}$\\
\hline
500 & 1.25 & 2.35 & 4.849 & 0.288 & 1.433 & 4.467 & 1.25 & 2.35 & 5.981 & 0.342 & 1.335 & 7.079 \\
1000 & 1.25 & 2.35 & 4.849 & 0.055 & 1.386 & 4.467 & 1.25 & 2.35 & 5.981 & 0.322 & 1.421 & 4.467 \\
\hline
 &\multicolumn{6}{c}{$z=6.8$}\\
 \hline
$Mecl_{min}$&$\alpha_{1,IMF}$&$\alpha_{2,IMF}$&$m_{c}$&$\alpha_{1,IGIMF}$&$\alpha_{2,IGIMF}$&$m_{c^{'}}$
\\
\hline
500 & 1.25 & 2.35 & 7.189 & 0.345 & 1.423 & 7.079 &  &  &  &  &  &  \\
1000 & 1.25 & 2.35 & 7.189 & 0.324 & 1.446 & 7.079 &  &  &  &  &  &  \\
\hline
\end{tabular*}
\end{tiny}
\end{table}
\clearpage

\begin{table}
\begin{tiny}

\caption{IGIMF and IMF slopes with varying z and $Mecl_{min}$ at
$\beta$=2,2.4 for SFR2:}
 \label{tab:final2}
\begin{tabular*}{7.055in}{ccccccccccccc}
\hline
\multicolumn{13}{c}{\textbf{$\beta=2.0$}}\\
 \hline
 &\multicolumn{6}{c}{$z=0.1$}&\multicolumn{6}{c}{$z=0.2$}\\
 \hline
$Mecl_{min}$&$\alpha_{1,IMF}$&$\alpha_{2,IMF}$&$m_{c}$&$\alpha_{1,IGIMF}$&$\alpha_{2,IGIMF}$&$m_{c^{'}}$
&$\alpha_{1,IMF}$&$\alpha_{2,IMF}$&$m_{c}$&$\alpha_{1,IGIMF}$&$\alpha_{2,IGIMF}$&$m_{c^{'}}$\\
\hline
500 & 1.25 & 2.35 & 0.381 & 0.429 & 1.415 & 0.224 & 1.25 & 2.35 & 0.434 & 0.019 & 1.437 & 0.282 \\
1000 & 1.25 & 2.35 & 0.381 & 0.312 & 1.321 & 0.224 & 1.25 & 2.35 & 0.434 & -0.054 & 1.404 & 0.282 \\
 \hline
 &\multicolumn{6}{c}{$z=0.8$}&\multicolumn{6}{c}{$z=1.5$}\\
 \hline
$Mecl_{min}$&$\alpha_{1,IMF}$&$\alpha_{2,IMF}$&$m_{c}$&$\alpha_{1,IGIMF}$&$\alpha_{2,IGIMF}$&$m_{c^{'}}$
&$\alpha_{1,IMF}$&$\alpha_{2,IMF}$&$m_{c}$&$\alpha_{1,IGIMF}$&$\alpha_{2,IGIMF}$&$m_{c^{'}}$\\
\hline
500 & 1.25 & 2.35 & 0.797 & 0.129 & 1.351 & 0.562 & 1.25 & 2.35 & 1.304 & 0.307 & 1.389 & 0.891 \\
1000 & 1.25 & 2.35 & 0.797 & 0.293 & 1.408 & 0.562 & 1.25 & 2.35 & 1.304 & 0.073 & 1.488 & 0.708 \\
\hline
 &\multicolumn{6}{c}{$z=2.2$}&\multicolumn{6}{c}{$z=3.8$}\\
 \hline
$Mecl_{min}$&$\alpha_{1,IMF}$&$\alpha_{2,IMF}$&$m_{c}$&$\alpha_{1,IGIMF}$&$\alpha_{2,IGIMF}$&$m_{c^{'}}$
&$\alpha_{1,IMF}$&$\alpha_{2,IMF}$&$m_{c}$&$\alpha_{1,IGIMF}$&$\alpha_{2,IGIMF}$&$m_{c^{'}}$\\
\hline
500 & 1.25 & 2.35 & 1.889 & 0.129 & 1.364 & 1.122 & 1.25 & 2.35 & 3.471 & 0.449 & 1.174 & 2.818 \\
1000 & 1.25 & 2.35 & 1.889 & 0.046 & 1.245 & 1.412 & 1.25 & 2.35 & 3.471 & 0.099 & 1.214 & 1.778 \\
\hline
 &\multicolumn{6}{c}{$z=5.0$}&\multicolumn{6}{c}{$z=5.9$}\\
 \hline
$Mecl_{min}$&$\alpha_{1,IMF}$&$\alpha_{2,IMF}$&$m_{c}$&$\alpha_{1,IGIMF}$&$\alpha_{2,IGIMF}$&$m_{c^{'}}$
&$\alpha_{1,IMF}$&$\alpha_{2,IMF}$&$m_{c}$&$\alpha_{1,IGIMF}$&$\alpha_{2,IGIMF}$&$m_{c^{'}}$\\
\hline
500 & 1.25 & 2.35 & 4.849 & 0.249 & 1.351 & 3.548 & 1.25 & 2.35 & 5.981 & 0.323 & 1.398 & 7.079 \\
1000 & 1.25 & 2.35 & 4.849 & 0.348 & 1.298 & 3.548 & 1.25 & 2.35 & 5.981 & 0.137 & 1.392 & 5.623 \\
 \hline
 &\multicolumn{6}{c}{$z=6.8$}\\
 \hline
$Mecl_{min}$&$\alpha_{1,IMF}$&$\alpha_{2,IMF}$&$m_{c}$&$\alpha_{1,IGIMF}$&$\alpha_{2,IGIMF}$&$m_{c^{'}}$
\\
\hline
500 & 1.25 & 2.35 & 7.189 & 0.241 & 1.202 & 7.079 &  &  &  &  &  &  \\
1000 & 1.25 & 2.35 & 7.189 & 0.377 & 1.223 & 7.079 &  &  &  &  &  &  \\
\hline
\multicolumn{13}{c}{\textbf{$\beta=2.4$}}\\
 \hline
 &\multicolumn{6}{c}{$z=0.1$}&\multicolumn{6}{c}{$z=0.2$}\\
 \hline
$Mecl_{min}$&$\alpha_{1,IMF}$&$\alpha_{2,IMF}$&$m_{c}$&$\alpha_{1,IGIMF}$&$\alpha_{2,IGIMF}$&$m_{c^{'}}$
&$\alpha_{1,IMF}$&$\alpha_{2,IMF}$&$m_{c}$&$\alpha_{1,IGIMF}$&$\alpha_{2,IGIMF}$&$m_{c^{'}}$\\
\hline
500 & 1.25 & 2.35 & 0.381 & 0.306 & 1.404 & 0.224 & 1.25 & 2.35 & 0.434 & 0.150 & 1.358 & 0.282 \\
1000 & 1.25 & 2.35 & 0.381 & 0.286 & 1.448 & 0.282 & 1.25 & 2.35 & 0.434 & 0.235 & 1.321 & 0.282 \\
\hline
 &\multicolumn{6}{c}{$z=0.8$}&\multicolumn{6}{c}{$z=1.5$}\\
 \hline
$Mecl_{min}$&$\alpha_{1,IMF}$&$\alpha_{2,IMF}$&$m_{c}$&$\alpha_{1,IGIMF}$&$\alpha_{2,IGIMF}$&$m_{c^{'}}$
&$\alpha_{1,IMF}$&$\alpha_{2,IMF}$&$m_{c}$&$\alpha_{1,IGIMF}$&$\alpha_{2,IGIMF}$&$m_{c^{'}}$\\
\hline
500 & 1.25 & 2.35 & 0.797 & 0.307 & 1.400 & 0.562 & 1.25 & 2.35 & 1.304 & 0.184 & 1.378 & 0.891 \\
1000 & 1.25 & 2.35 & 0.797 & 0.379 & 1.446 & 0.562 & 1.25 & 2.35 & 1.304 & 0.170 & 1.432 & 0.891 \\
\hline
 &\multicolumn{6}{c}{$z=2.2$}&\multicolumn{6}{c}{$z=3.8$}\\
 \hline
$Mecl_{min}$&$\alpha_{1,IMF}$&$\alpha_{2,IMF}$&$m_{c}$&$\alpha_{1,IGIMF}$&$\alpha_{2,IGIMF}$&$m_{c^{'}}$
&$\alpha_{1,IMF}$&$\alpha_{2,IMF}$&$m_{c}$&$\alpha_{1,IGIMF}$&$\alpha_{2,IGIMF}$&$m_{c^{'}}$\\
\hline
500 & 1.25 & 2.35 & 1.889 & 0.383 & 1.365 & 1.778 & 1.25 & 2.35 & 3.471 & 0.186 & 1.351 & 2.818 \\
1000 & 1.25 & 2.35 & 1.889 & 0.253 & 1.392 & 1.412 & 1.25 & 2.35 & 3.471 & 0.027 & 1.264 & 2.238 \\
 \hline
 &\multicolumn{6}{c}{$z=5.0$}&\multicolumn{6}{c}{$z=5.9$}\\
 \hline
$Mecl_{min}$&$\alpha_{1,IMF}$&$\alpha_{2,IMF}$&$m_{c}$&$\alpha_{1,IGIMF}$&$\alpha_{2,IGIMF}$&$m_{c^{'}}$
&$\alpha_{1,IMF}$&$\alpha_{2,IMF}$&$m_{c}$&$\alpha_{1,IGIMF}$&$\alpha_{2,IGIMF}$&$m_{c^{'}}$\\
\hline
500 & 1.25 & 2.35 & 4.849 & 0.304 & 1.420 & 3.548 & 1.25 & 2.35 & 5.981 & 0.335 & 1.333 & 5.623 \\
1000 & 1.25 & 2.35 & 4.849 & 0.167 & 1.425 & 5.623 & 1.25 & 2.35 & 5.981 & 0.313 & 1.495 & 5.623 \\
\hline
 &\multicolumn{6}{c}{$z=6.8$}\\
 \hline
$Mecl_{min}$&$\alpha_{1,IMF}$&$\alpha_{2,IMF}$&$m_{c}$&$\alpha_{1,IGIMF}$&$\alpha_{2,IGIMF}$&$m_{c^{'}}$
\\
\hline
500 & 1.25 & 2.35 & 7.189 & 0.219 & 1.320 & 5.623 &  &  &  &  &  &  \\
1000 & 1.25 & 2.35 & 7.189 & 0.167 & 1.433 & 5.623 &  &  &  &  &  &  \\
\hline
\end{tabular*}
\end{tiny}
\end{table}

\begin{table}
\begin{tiny}

\caption{IGIMF and IMF slopes with varying z and $Mecl_{min}$ at
$\beta$=2.6 for SFR2:}
 \label{tab:final2}
\begin{tabular*}{7.055in}{ccccccccccccc}
\hline
\multicolumn{13}{c}{\textbf{$\beta=2.6$}}\\
 \hline
 &\multicolumn{6}{c}{$z=0.1$}&\multicolumn{6}{c}{$z=0.2$}\\
 \hline
$Mecl_{min}$&$\alpha_{1,IMF}$&$\alpha_{2,IMF}$&$m_{c}$&$\alpha_{1,IGIMF}$&$\alpha_{2,IGIMF}$&$m_{c^{'}}$
&$\alpha_{1,IMF}$&$\alpha_{2,IMF}$&$m_{c}$&$\alpha_{1,IGIMF}$&$\alpha_{2,IGIMF}$&$m_{c^{'}}$\\
\hline
500 & 1.25 & 2.35 & 0.381 & 0.388 & 1.433 & 0.282 & 1.25 & 2.35 & 0.434 & -0.404 & 1.400 & 0.224 \\
1000 & 1.25 & 2.35 & 0.381 & 0.195 & 1.456 & 0.224 & 1.25 & 2.35 & 0.434 & 0.645 & 1.448 & 0.282 \\
\hline
 &\multicolumn{6}{c}{$z=0.8$}&\multicolumn{6}{c}{$z=1.5$}\\
 \hline
$Mecl_{min}$&$\alpha_{1,IMF}$&$\alpha_{2,IMF}$&$m_{c}$&$\alpha_{1,IGIMF}$&$\alpha_{2,IGIMF}$&$m_{c^{'}}$
&$\alpha_{1,IMF}$&$\alpha_{2,IMF}$&$m_{c}$&$\alpha_{1,IGIMF}$&$\alpha_{2,IGIMF}$&$m_{c^{'}}$\\
\hline
500 & 1.25 & 2.35 & 0.797 & 0.213 & 1.417 & 0.562 & 1.25 & 2.35 & 1.304 & 0.212 & 1.371 & 1.122 \\
1000 & 1.25 & 2.35 & 0.797 & -0.049 & 1.403 & 0.708 & 1.25 & 2.35 & 1.304 & 0.291 & 1.446 & 1.122 \\
 \hline
 &\multicolumn{6}{c}{$z=2.2$}&\multicolumn{6}{c}{$z=3.8$}\\
 \hline
$Mecl_{min}$&$\alpha_{1,IMF}$&$\alpha_{2,IMF}$&$m_{c}$&$\alpha_{1,IGIMF}$&$\alpha_{2,IGIMF}$&$m_{c^{'}}$
&$\alpha_{1,IMF}$&$\alpha_{2,IMF}$&$m_{c}$&$\alpha_{1,IGIMF}$&$\alpha_{2,IGIMF}$&$m_{c^{'}}$\\
\hline
500 & 1.25 & 2.35 & 1.889 & 0.173 & 1.413 & 1.412 & 1.25 & 2.35 & 3.471 & 0.284 & 1.406 & 1.412 \\
1000 & 1.25 & 2.35 & 1.889 & 0.298 & 1.394 & 1.412 & 1.25 & 2.35 & 3.471 & 0.221 & 1.333 & 2.818 \\
\hline
 &\multicolumn{6}{c}{$z=5.0$}&\multicolumn{6}{c}{$z=5.9$}\\
 \hline
$Mecl_{min}$&$\alpha_{1,IMF}$&$\alpha_{2,IMF}$&$m_{c}$&$\alpha_{1,IGIMF}$&$\alpha_{2,IGIMF}$&$m_{c^{'}}$
&$\alpha_{1,IMF}$&$\alpha_{2,IMF}$&$m_{c}$&$\alpha_{1,IGIMF}$&$\alpha_{2,IGIMF}$&$m_{c^{'}}$\\
\hline
500 & 1.25 & 2.35 & 4.849 & 0.218 & 1.412 & 4.467 & 1.25 & 2.35 & 5.981 & 0.241 & 1.385 & 5.623 \\
1000 & 1.25 & 2.35 & 4.849 & 0.207 & 1.333 & 3.548 & 1.25 & 2.35 & 5.981 & 0.329 & 1.264 & 4.467 \\
\hline
 &\multicolumn{6}{c}{$z=6.8$}\\
 \hline
$Mecl_{min}$&$\alpha_{1,IMF}$&$\alpha_{2,IMF}$&$m_{c}$&$\alpha_{1,IGIMF}$&$\alpha_{2,IGIMF}$&$m_{c^{'}}$
\\
\hline
500 & 1.25 & 2.35 & 7.189 & 0.273 & 1.391 & 7.079 &  &  &  &  &  &  \\
1000 & 1.25 & 2.35 & 7.189 & 0.238 & 1.283 & 7.079 &  &  &  &  &  &  \\
\hline
\end{tabular*}
\end{tiny}
\end{table}

\clearpage
\begin{table}
\begin{tiny}

\caption{IGIMF and IMF slopes with varying z and $Mecl_{min}$ at
$\beta$=2,2.4 for SFR3:}
 \label{tab:final2}
\begin{tabular*}{7.055in}{ccccccccccccc}
\hline
\multicolumn{13}{c}{\textbf{$\beta=2.0$}}\\
 \hline
 &\multicolumn{6}{c}{$z=0.1$}&\multicolumn{6}{c}{$z=0.2$}\\
 \hline
$Mecl_{min}$&$\alpha_{1,IMF}$&$\alpha_{2,IMF}$&$m_{c}$&$\alpha_{1,IGIMF}$&$\alpha_{2,IGIMF}$&$m_{c^{'}}$
&$\alpha_{1,IMF}$&$\alpha_{2,IMF}$&$m_{c}$&$\alpha_{1,IGIMF}$&$\alpha_{2,IGIMF}$&$m_{c^{'}}$\\
\hline
500 & 1.25 & 2.35 & 0.381 & 0.428 & 1.369 & 0.282 & 1.25 & 2.35 & 0.434 & 0.008 & 1.428 & 0.282 \\
1000 & 1.25 & 2.35 & 0.381 & -0.795 & 1.414 & 0.178 & 1.25 & 2.35 & 0.434 & -0.236 & 1.429 & 0.282 \\
 \hline
 &\multicolumn{6}{c}{$z=0.8$}&\multicolumn{6}{c}{$z=1.5$}\\
 \hline
$Mecl_{min}$&$\alpha_{1,IMF}$&$\alpha_{2,IMF}$&$m_{c}$&$\alpha_{1,IGIMF}$&$\alpha_{2,IGIMF}$&$m_{c^{'}}$
&$\alpha_{1,IMF}$&$\alpha_{2,IMF}$&$m_{c}$&$\alpha_{1,IGIMF}$&$\alpha_{2,IGIMF}$&$m_{c^{'}}$\\
\hline
500 & 1.25 & 2.35 & 0.797 & 0.188 & 1.316 & 0.562 & 1.25 & 2.35 & 1.304 & 0.134 & 1.401 & 0.562 \\
1000 & 1.25 & 2.35 & 0.797 & 0.276 & 1.485 & 0.447 & 1.25 & 2.35 & 1.304 & 0.279 & 1.302 & 1.122 \\
\hline
 &\multicolumn{6}{c}{$z=2.2$}&\multicolumn{6}{c}{$z=3.8$}\\
 \hline
$Mecl_{min}$&$\alpha_{1,IMF}$&$\alpha_{2,IMF}$&$m_{c}$&$\alpha_{1,IGIMF}$&$\alpha_{2,IGIMF}$&$m_{c^{'}}$
&$\alpha_{1,IMF}$&$\alpha_{2,IMF}$&$m_{c}$&$\alpha_{1,IGIMF}$&$\alpha_{2,IGIMF}$&$m_{c^{'}}$\\
\hline
500 & 1.25 & 2.35 & 1.889 & 0.203 & 1.421 & 1.412 & 1.25 & 2.35 & 3.471 & -0.117 & 1.232 & 3.548 \\
1000 & 1.25 & 2.35 & 1.889 & -0.072 & 1.314 & 1.412 & 1.25 & 2.35 & 3.471 & 0.273 & 1.512 & 2.818 \\
\hline
 &\multicolumn{6}{c}{$z=5.0$}&\multicolumn{6}{c}{$z=5.9$}\\
 \hline
$Mecl_{min}$&$\alpha_{1,IMF}$&$\alpha_{2,IMF}$&$m_{c}$&$\alpha_{1,IGIMF}$&$\alpha_{2,IGIMF}$&$m_{c^{'}}$
&$\alpha_{1,IMF}$&$\alpha_{2,IMF}$&$m_{c}$&$\alpha_{1,IGIMF}$&$\alpha_{2,IGIMF}$&$m_{c^{'}}$\\
\hline
500 & 1.25 & 2.35 & 4.849 & 0.269 & 1.385 & 3.548 & 1.25 & 2.35 & 5.981 & 0.313 & 1.398 & 7.079 \\
1000 & 1.25 & 2.35 & 4.849 & 0.382 & 1.376 & 3.548 & 1.25 & 2.35 & 5.981 & 0.446 & 1.375 & 5.623 \\
 \hline
 &\multicolumn{6}{c}{$z=6.8$}\\
 \hline
$Mecl_{min}$&$\alpha_{1,IMF}$&$\alpha_{2,IMF}$&$m_{c}$&$\alpha_{1,IGIMF}$&$\alpha_{2,IGIMF}$&$m_{c^{'}}$
\\
\hline
500 & 1.25 & 2.35 & 7.189 & 0.178 & 1.497 & 7.079 &  &  &  &  &  &  \\
1000 & 1.25 & 2.35 & 7.189 & 0.419 & 1.325 & 7.079 &  &  &  &  &  &  \\
\hline
\multicolumn{13}{c}{\textbf{$\beta=2.4$}}\\
 \hline
 &\multicolumn{6}{c}{$z=0.1$}&\multicolumn{6}{c}{$z=0.2$}\\
 \hline
$Mecl_{min}$&$\alpha_{1,IMF}$&$\alpha_{2,IMF}$&$m_{c}$&$\alpha_{1,IGIMF}$&$\alpha_{2,IGIMF}$&$m_{c^{'}}$
&$\alpha_{1,IMF}$&$\alpha_{2,IMF}$&$m_{c}$&$\alpha_{1,IGIMF}$&$\alpha_{2,IGIMF}$&$m_{c^{'}}$\\
\hline
500 & 1.25 & 2.35 & 0.381 & 0.009 & 1.491 & 0.224 & 1.25 & 2.35 & 0.434 & 0.343 & 1.393 & 0.282 \\
1000 & 1.25 & 2.35 & 0.381 & 0.395 & 1.416 & 0.224 & 1.25 & 2.35 & 0.434 & 0.023 & 1.362 & 0.282 \\
\hline
 &\multicolumn{6}{c}{$z=0.8$}&\multicolumn{6}{c}{$z=1.5$}\\
 \hline
$Mecl_{min}$&$\alpha_{1,IMF}$&$\alpha_{2,IMF}$&$m_{c}$&$\alpha_{1,IGIMF}$&$\alpha_{2,IGIMF}$&$m_{c^{'}}$
&$\alpha_{1,IMF}$&$\alpha_{2,IMF}$&$m_{c}$&$\alpha_{1,IGIMF}$&$\alpha_{2,IGIMF}$&$m_{c^{'}}$\\
\hline
500 & 1.25 & 2.35 & 0.797 & 0.315 & 1.349 & 0.447 & 1.25 & 2.35 & 1.304 & 0.226 & 1.386 & 0.708 \\
1000 & 1.25 & 2.35 & 0.797 & 0.321 & 1.407 & 0.562 & 1.25 & 2.35 & 1.304 & 0.133 & 1.501 & 1.122 \\
\hline
 &\multicolumn{6}{c}{$z=2.2$}&\multicolumn{6}{c}{$z=3.8$}\\
 \hline
$Mecl_{min}$&$\alpha_{1,IMF}$&$\alpha_{2,IMF}$&$m_{c}$&$\alpha_{1,IGIMF}$&$\alpha_{2,IGIMF}$&$m_{c^{'}}$
&$\alpha_{1,IMF}$&$\alpha_{2,IMF}$&$m_{c}$&$\alpha_{1,IGIMF}$&$\alpha_{2,IGIMF}$&$m_{c^{'}}$\\
\hline
500 & 1.25 & 2.35 & 1.889 & 0.227 & 1.361 & 1.412 & 1.25 & 2.35 & 3.471 & 0.130 & 1.315 & 2.818 \\
1000 & 1.25 & 2.35 & 1.889 & 0.462 & 1.380 & 1.778 & 1.25 & 2.35 & 3.471 & 0.076 & 1.300 & 3.548 \\
 \hline
 &\multicolumn{6}{c}{$z=5.0$}&\multicolumn{6}{c}{$z=5.9$}\\
 \hline
$Mecl_{min}$&$\alpha_{1,IMF}$&$\alpha_{2,IMF}$&$m_{c}$&$\alpha_{1,IGIMF}$&$\alpha_{2,IGIMF}$&$m_{c^{'}}$
&$\alpha_{1,IMF}$&$\alpha_{2,IMF}$&$m_{c}$&$\alpha_{1,IGIMF}$&$\alpha_{2,IGIMF}$&$m_{c^{'}}$\\
\hline
500 & 1.25 & 2.35 & 4.849 & 0.219 & 1.372 & 3.548 & 1.25 & 2.35 & 5.981 & 0.215 & 1.322 & 5.623 \\
1000 & 1.25 & 2.35 & 4.849 & -0.076 & 1.315 & 4.467 & 1.25 & 2.35 & 5.981 & 0.131 & 1.367 & 5.623 \\
\hline
 &\multicolumn{6}{c}{$z=6.8$}\\
 \hline
$Mecl_{min}$&$\alpha_{1,IMF}$&$\alpha_{2,IMF}$&$m_{c}$&$\alpha_{1,IGIMF}$&$\alpha_{2,IGIMF}$&$m_{c^{'}}$
\\
\hline
500 & 1.25 & 2.35 & 7.189 & 0.337 & 1.420 & 5.623 &  &  &  &  &  &  \\
1000 & 1.25 & 2.35 & 7.189 & 0.268 & 1.349 & 7.079 &  &  &  &  &  &  \\
\hline
\end{tabular*}
\end{tiny}
\end{table}
\clearpage

\begin{table}
\begin{tiny}

\caption{IGIMF and IMF slopes with varying z and $Mecl_{min}$ at
$\beta$=2.6 for SFR3:}
 \label{tab:final2}
\begin{tabular*}{7.055in}{ccccccccccccc}
\hline
\multicolumn{13}{c}{\textbf{$\beta=2.6$}}\\
 \hline
 &\multicolumn{6}{c}{$z=0.1$}&\multicolumn{6}{c}{$z=0.2$}\\
 \hline
$Mecl_{min}$&$\alpha_{1,IMF}$&$\alpha_{2,IMF}$&$m_{c}$&$\alpha_{1,IGIMF}$&$\alpha_{2,IGIMF}$&$m_{c^{'}}$
&$\alpha_{1,IMF}$&$\alpha_{2,IMF}$&$m_{c}$&$\alpha_{1,IGIMF}$&$\alpha_{2,IGIMF}$&$m_{c^{'}}$\\
\hline
500 & 1.25 & 2.35 & 0.381 & -0.313 & 1.391 & 0.224 & 1.25 & 2.35 & 0.434 & 0.325 & 1.419 & 0.282 \\
1000 & 1.25 & 2.35 & 0.381 & -0.017 & 1.393 & 0.224 & 1.25 & 2.35 & 0.434 & 0.052 & 1.500 & 0.282 \\
\hline
 &\multicolumn{6}{c}{$z=0.8$}&\multicolumn{6}{c}{$z=1.5$}\\
 \hline
$Mecl_{min}$&$\alpha_{1,IMF}$&$\alpha_{2,IMF}$&$m_{c}$&$\alpha_{1,IGIMF}$&$\alpha_{2,IGIMF}$&$m_{c^{'}}$
&$\alpha_{1,IMF}$&$\alpha_{2,IMF}$&$m_{c}$&$\alpha_{1,IGIMF}$&$\alpha_{2,IGIMF}$&$m_{c^{'}}$\\
\hline
500 & 1.25 & 2.35 & 0.797 & 0.402 & 1.462 & 0.562 & 1.25 & 2.35 & 1.304 & 0.383 & 1.444 & 0.891 \\
1000 & 1.25 & 2.35 & 0.797 & 0.385 & 1.488 & 0.562 & 1.25 & 2.35 & 1.304 & 0.358 & 1.384 & 0.891 \\
 \hline
 &\multicolumn{6}{c}{$z=2.2$}&\multicolumn{6}{c}{$z=3.8$}\\
 \hline
$Mecl_{min}$&$\alpha_{1,IMF}$&$\alpha_{2,IMF}$&$m_{c}$&$\alpha_{1,IGIMF}$&$\alpha_{2,IGIMF}$&$m_{c^{'}}$
&$\alpha_{1,IMF}$&$\alpha_{2,IMF}$&$m_{c}$&$\alpha_{1,IGIMF}$&$\alpha_{2,IGIMF}$&$m_{c^{'}}$\\
\hline
500 & 1.25 & 2.35 & 1.889 & 0.246 & 1.366 & 1.412 & 1.25 & 2.35 & 3.471 & 0.266 & 1.328 & 2.818 \\
1000 & 1.25 & 2.35 & 1.889 & 0.307 & 1.370 & 0.891 & 1.25 & 2.35 & 3.471 & 0.294 & 1.426 & 2.818 \\
\hline
 &\multicolumn{6}{c}{$z=5.0$}&\multicolumn{6}{c}{$z=5.9$}\\
 \hline
$Mecl_{min}$&$\alpha_{1,IMF}$&$\alpha_{2,IMF}$&$m_{c}$&$\alpha_{1,IGIMF}$&$\alpha_{2,IGIMF}$&$m_{c^{'}}$
&$\alpha_{1,IMF}$&$\alpha_{2,IMF}$&$m_{c}$&$\alpha_{1,IGIMF}$&$\alpha_{2,IGIMF}$&$m_{c^{'}}$\\
\hline
500 & 1.25 & 2.35 & 4.849 & 0.337 & 1.342 & 4.467 & 1.25 & 2.35 & 5.981 & 0.295 & 1.351 & 5.623 \\
1000 & 1.25 & 2.35 & 4.849 & 0.258 & 1.351 & 3.548 & 1.25 & 2.35 & 5.981 & 0.306 & 1.409 & 5.623 \\
\hline
 &\multicolumn{6}{c}{$z=6.8$}\\
 \hline
$Mecl_{min}$&$\alpha_{1,IMF}$&$\alpha_{2,IMF}$&$m_{c}$&$\alpha_{1,IGIMF}$&$\alpha_{2,IGIMF}$&$m_{c^{'}}$
\\
\hline
500 & 1.25 & 2.35 & 7.189 & 0.279 & 1.381 & 5.623 &  &  &  &  &  &  \\
1000 & 1.25 & 2.35 & 7.189 & 0.187 & 1.506 & 7.079 &  &  &  &  &  &  \\
\hline
\end{tabular*}
\end{tiny}
\end{table}
\clearpage

\begin{table}
\begin{tiny}

\caption{IGIMF and IMF slopes with varying z and $Mecl_{min}$ at
$\beta$=2,2.4 for $SFR^{*}$:}
 \label{tab:final2}
\begin{tabular*}{7.055in}{ccccccccccccc}
\hline
\multicolumn{13}{c}{\textbf{$\beta=2.0$}}\\
 \hline
 &\multicolumn{6}{c}{$z=0.1$}&\multicolumn{6}{c}{$z=0.2$}\\
 \hline
$Mecl_{min}$&$\alpha_{1,IMF}$&$\alpha_{2,IMF}$&$m_{c}$&$\alpha_{1,IGIMF}$&$\alpha_{2,IGIMF}$&$m_{c^{'}}$
&$\alpha_{1,IMF}$&$\alpha_{2,IMF}$&$m_{c}$&$\alpha_{1,IGIMF}$&$\alpha_{2,IGIMF}$&$m_{c^{'}}$\\
\hline
500 & 1.25 & 2.35 & 0.381 & -0.265 & 1.478 & 0.224 & 1.25 & 2.35 & 0.434 & 0.532 & 1.371 & 0.224 \\
1000 & 1.25 & 2.35 & 0.381 & 0.644 & 1.375 & 0.224 & 1.25 & 2.35 & 0.434 & 0.354 & 1.362 & 0.224 \\
 \hline
 &\multicolumn{6}{c}{$z=0.8$}&\multicolumn{6}{c}{$z=1.5$}\\
 \hline
$Mecl_{min}$&$\alpha_{1,IMF}$&$\alpha_{2,IMF}$&$m_{c}$&$\alpha_{1,IGIMF}$&$\alpha_{2,IGIMF}$&$m_{c^{'}}$
&$\alpha_{1,IMF}$&$\alpha_{2,IMF}$&$m_{c}$&$\alpha_{1,IGIMF}$&$\alpha_{2,IGIMF}$&$m_{c^{'}}$\\
\hline
500 & 1.25 & 2.35 & 0.797 & 0.050 & 1.502 & 0.708 & 1.25 & 2.35 & 1.304 & 0.055 & 1.448 & 0.891 \\
1000 & 1.25 & 2.35 & 0.797 & 0.494 & 1.449 & 0.708 & 1.25 & 2.35 & 1.304 & 0.093 & 1.242 & 0.891 \\
\hline
 &\multicolumn{6}{c}{$z=2.2$}&\multicolumn{6}{c}{$z=3.8$}\\
 \hline
$Mecl_{min}$&$\alpha_{1,IMF}$&$\alpha_{2,IMF}$&$m_{c}$&$\alpha_{1,IGIMF}$&$\alpha_{2,IGIMF}$&$m_{c^{'}}$
&$\alpha_{1,IMF}$&$\alpha_{2,IMF}$&$m_{c}$&$\alpha_{1,IGIMF}$&$\alpha_{2,IGIMF}$&$m_{c^{'}}$\\
\hline
500 & 1.25 & 2.35 & 1.889 & 0.021 & 1.527 & 1.778 & 1.25 & 2.35 & 3.471 & 0.121 & 1.368 & 1.778 \\
1000 & 1.25 & 2.35 & 1.889 & 0.256 & 1.465 & 1.412 & 1.25 & 2.35 & 3.471 & 0.318 & 1.395 & 2.239 \\
\hline
 &\multicolumn{6}{c}{$z=5.0$}&\multicolumn{6}{c}{$z=5.9$}\\
 \hline
$Mecl_{min}$&$\alpha_{1,IMF}$&$\alpha_{2,IMF}$&$m_{c}$&$\alpha_{1,IGIMF}$&$\alpha_{2,IGIMF}$&$m_{c^{'}}$
&$\alpha_{1,IMF}$&$\alpha_{2,IMF}$&$m_{c}$&$\alpha_{1,IGIMF}$&$\alpha_{2,IGIMF}$&$m_{c^{'}}$\\
\hline
500 & 1.25 & 2.35 & 4.849 & 0.117 & 1.359 & 4.467 & 1.25 & 2.35 & 5.981 & 0.152 & 1.347 & 4.467 \\
1000 & 1.25 & 2.35 & 4.849 & 0.472 & 1.335 & 4.467 & 1.25 & 2.35 & 5.981 & 0..373 & 1.368 & 4.467 \\
 \hline
 &\multicolumn{6}{c}{$z=6.8$}\\
 \hline
$Mecl_{min}$&$\alpha_{1,IMF}$&$\alpha_{2,IMF}$&$m_{c}$&$\alpha_{1,IGIMF}$&$\alpha_{2,IGIMF}$&$m_{c^{'}}$
\\
\hline
500 & 1.25 & 2.35 & 7.189 & 0.215 & 1.381 & 7.079 &  &  &  &  &  &  \\
1000 & 1.25 & 2.35 & 7.189 & 0.145 & 1.564 & 7.079 &  &  &  &  &  &  \\
\hline
\multicolumn{13}{c}{\textbf{$\beta=2.4$}}\\
 \hline
 &\multicolumn{6}{c}{$z=0.1$}&\multicolumn{6}{c}{$z=0.2$}\\
 \hline
$Mecl_{min}$&$\alpha_{1,IMF}$&$\alpha_{2,IMF}$&$m_{c}$&$\alpha_{1,IGIMF}$&$\alpha_{2,IGIMF}$&$m_{c^{'}}$
&$\alpha_{1,IMF}$&$\alpha_{2,IMF}$&$m_{c}$&$\alpha_{1,IGIMF}$&$\alpha_{2,IGIMF}$&$m_{c^{'}}$\\
\hline
500 & 1.25 & 2.35 & 0.381 & 0.115 & 1.474 & 0.224 & 1.25 & 2.35 & 0.434 & 0.439 & 1.409 & 0.282 \\
1000 & 1.25 & 2.35 & 0.381 & -0.037 & 1.415 & 0.224 & 1.25 & 2.35 & 0.434 & 0.529 & 1.363 & 0.282 \\
\hline
 &\multicolumn{6}{c}{$z=0.8$}&\multicolumn{6}{c}{$z=1.5$}\\
 \hline
$Mecl_{min}$&$\alpha_{1,IMF}$&$\alpha_{2,IMF}$&$m_{c}$&$\alpha_{1,IGIMF}$&$\alpha_{2,IGIMF}$&$m_{c^{'}}$
&$\alpha_{1,IMF}$&$\alpha_{2,IMF}$&$m_{c}$&$\alpha_{1,IGIMF}$&$\alpha_{2,IGIMF}$&$m_{c^{'}}$\\
\hline
500 & 1.25 & 2.35 & 0.797 & 0.221 & 1.448 & 0.447 & 1.25 & 2.35 & 1.304 & 0.331 & 1.381 & 0.891 \\
1000 & 1.25 & 2.35 & 0.797 & 0.076 & 1.412 & 0.562 & 1.25 & 2.35 & 1.304 & 0.218 & 1.446 & 0.891 \\
\hline
 &\multicolumn{6}{c}{$z=2.2$}&\multicolumn{6}{c}{$z=3.8$}\\
 \hline
$Mecl_{min}$&$\alpha_{1,IMF}$&$\alpha_{2,IMF}$&$m_{c}$&$\alpha_{1,IGIMF}$&$\alpha_{2,IGIMF}$&$m_{c^{'}}$
&$\alpha_{1,IMF}$&$\alpha_{2,IMF}$&$m_{c}$&$\alpha_{1,IGIMF}$&$\alpha_{2,IGIMF}$&$m_{c^{'}}$\\
\hline
500 & 1.25 & 2.35 & 1.889 & 0.135 & 1.307 & 1.778 & 1.25 & 2.35 & 3.471 & 0.210 & 1.340 & 2.818 \\
1000 & 1.25 & 2.35 & 1.889 & 0.329 & 1.332 & 1.412 & 1.25 & 2.35 & 3.471 & 0.191 & 1.412 & 3.548 \\
 \hline
 &\multicolumn{6}{c}{$z=5.0$}&\multicolumn{6}{c}{$z=5.9$}\\
 \hline
$Mecl_{min}$&$\alpha_{1,IMF}$&$\alpha_{2,IMF}$&$m_{c}$&$\alpha_{1,IGIMF}$&$\alpha_{2,IGIMF}$&$m_{c^{'}}$
&$\alpha_{1,IMF}$&$\alpha_{2,IMF}$&$m_{c}$&$\alpha_{1,IGIMF}$&$\alpha_{2,IGIMF}$&$m_{c^{'}}$\\
\hline
500 & 1.25 & 2.35 & 4.849 & 0.226 & 1.392 & 4.467 & 1.25 & 2.35 & 5.981 & -0.022 & 1.339 & 4.467 \\
1000 & 1.25 & 2.35 & 4.849 & 0.281 & 1.274 & 4.467 & 1.25 & 2.35 & 5.981 & 0.149 & 1.329 & 4.467 \\
\hline
 &\multicolumn{6}{c}{$z=6.8$}\\
 \hline
$Mecl_{min}$&$\alpha_{1,IMF}$&$\alpha_{2,IMF}$&$m_{c}$&$\alpha_{1,IGIMF}$&$\alpha_{2,IGIMF}$&$m_{c^{'}}$
\\
\hline
500 & 1.25 & 2.35 & 7.189 & 0.330 & 1.373 & 7.079 &  &  &  &  &  &  \\
1000 & 1.25 & 2.35 & 7.189 & 0.234 & 1.342 & 5.623 &  &  &  &  &  &  \\
\hline
\end{tabular*}
\end{tiny}
\end{table}

\clearpage

\begin{table}
\begin{tiny}

\caption{IGIMF and IMF slopes with varying z and $Mecl_{min}$ at
$\beta$=2.6 for $SFR^{*}$:}
 \label{tab:final2}
\begin{tabular*}{7.055in}{ccccccccccccc}
\hline
\multicolumn{13}{c}{\textbf{$\beta=2.6$}}\\
 \hline
 &\multicolumn{6}{c}{$z=0.1$}&\multicolumn{6}{c}{$z=0.2$}\\
 \hline
$Mecl_{min}$&$\alpha_{1,IMF}$&$\alpha_{2,IMF}$&$m_{c}$&$\alpha_{1,IGIMF}$&$\alpha_{2,IGIMF}$&$m_{c^{'}}$
&$\alpha_{1,IMF}$&$\alpha_{2,IMF}$&$m_{c}$&$\alpha_{1,IGIMF}$&$\alpha_{2,IGIMF}$&$m_{c^{'}}$\\
\hline
500 & 1.25 & 2.35 & 0.381 & 0.150 & 1.442 & 0.224 & 1.25 & 2.35 & 0.434 & 0.241 & 1.416 & 0.282 \\
1000 & 1.25 & 2.35 & 0.381 & 0.211 & 1.416 & 0.224 & 1.25 & 2.35 & 0.434 & 0.240 & 1.418 & 0.282 \\
\hline
 &\multicolumn{6}{c}{$z=0.8$}&\multicolumn{6}{c}{$z=1.5$}\\
 \hline
$Mecl_{min}$&$\alpha_{1,IMF}$&$\alpha_{2,IMF}$&$m_{c}$&$\alpha_{1,IGIMF}$&$\alpha_{2,IGIMF}$&$m_{c^{'}}$
&$\alpha_{1,IMF}$&$\alpha_{2,IMF}$&$m_{c}$&$\alpha_{1,IGIMF}$&$\alpha_{2,IGIMF}$&$m_{c^{'}}$\\
\hline
500 & 1.25 & 2.35 & 0.797 & 0.156 & 1.412 & 0.562 & 1.25 & 2.35 & 1.304 & 0.196 & 1.370 & 0.891 \\
1000 & 1.25 & 2.35 & 0.797 & 0.260 & 1.426 & 0.562 & 1.25 & 2.35 & 1.304 & 0.224 & 1.463 & 1.122 \\
 \hline
 &\multicolumn{6}{c}{$z=2.2$}&\multicolumn{6}{c}{$z=3.8$}\\
 \hline
$Mecl_{min}$&$\alpha_{1,IMF}$&$\alpha_{2,IMF}$&$m_{c}$&$\alpha_{1,IGIMF}$&$\alpha_{2,IGIMF}$&$m_{c^{'}}$
&$\alpha_{1,IMF}$&$\alpha_{2,IMF}$&$m_{c}$&$\alpha_{1,IGIMF}$&$\alpha_{2,IGIMF}$&$m_{c^{'}}$\\
\hline
500 & 1.25 & 2.35 & 1.889 & 0.366 & 1.378 & 1.412 & 1.25 & 2.35 & 3.471 & 0.238 & 1.484 & 3.548 \\
1000 & 1.25 & 2.35 & 1.889 & 0.295 & 1.487 & 1.412 & 1.25 & 2.35 & 3.471 & 0.276 & 1.261 & 2.818 \\
\hline
 &\multicolumn{6}{c}{$z=5.0$}&\multicolumn{6}{c}{$z=5.9$}\\
 \hline
$Mecl_{min}$&$\alpha_{1,IMF}$&$\alpha_{2,IMF}$&$m_{c}$&$\alpha_{1,IGIMF}$&$\alpha_{2,IGIMF}$&$m_{c^{'}}$
&$\alpha_{1,IMF}$&$\alpha_{2,IMF}$&$m_{c}$&$\alpha_{1,IGIMF}$&$\alpha_{2,IGIMF}$&$m_{c^{'}}$\\
\hline
500 & 1.25 & 2.35 & 4.849 & 0.319 & 1.258 & 4.467 & 1.25 & 2.35 & 5.981 & 0.299 & 1.339 & 4.467 \\
1000 & 1.25 & 2.35 & 4.849 & 0.344 & 1.523 & 5.623 & 1.25 & 2.35 & 5.981 & 0.314 & 1.281 & 5.623 \\
\hline
 &\multicolumn{6}{c}{$z=6.8$}\\
 \hline
$Mecl_{min}$&$\alpha_{1,IMF}$&$\alpha_{2,IMF}$&$m_{c}$&$\alpha_{1,IGIMF}$&$\alpha_{2,IGIMF}$&$m_{c^{'}}$
\\
\hline
500 & 1.25 & 2.35 & 7.189 & 0.158 & 1.512 & 7.079 &  &  &  &  &  &  \\
1000 & 1.25 & 2.35 & 7.189 & 0.192 & 1.402 & 7.079 &  &  &  &  &  &  \\
\hline
\end{tabular*}
\end{tiny}
\end{table}

\clearpage

\end{document}